\documentclass[sn-mathphys]{sn-jnl}
\usepackage{amssymb,mathtools,bm,mathrsfs}

\newcommand{\vD}{\scalebox{0.8}[1]{$\varDelta$}}
\newcommand{\vg}{\text{\usefont{OML}{cmm}{m}{it}g}}

\jyear{2024}%

\begin{document}

\title[Structure of Point Particles]
{On Geometric Structure of Point Particles}

\author*[]{\fnm{M.} \sur{Honda}}

\affil{\orgdiv{Plasma Astrophysics Laboratory},
\orgname{Institute for Global Science},
\orgaddress{\state{Mie}, \country{Japan}}}

\abstract{I propose, as geometric structure in an internal space,
a helical field that is responsible for intrinsic properties of
point particles, particularly, electron.
For the novel theoretical development, plasma astrophysical analogy
is made extensively.
Transition between our conventional space and the infinitesimal space
is considered in an operational manner.
It is shown that rotational eigenvalue equation satisfied by the vector
field equivalent to Gromeka-Beltrami flow provides a coordinate-rotor
that captures complex orthogonality between the internal
coordinate space and isotopic, angular momentum space.
Self-consistent normalization of rotational coordinate owing to the
rotor is compared to renormalization of electric charge.
It is also found that chiral asymmetry of the helical eigenflows can be
reflected in electroweak symmetry breaking.
The theoretical prototype suggests possible geometrical features of a
fundamental framework ruling matters and forces with higher dimensions.}


\maketitle

\section{Introduction}\label{sec:1}

Reducing elements of matters has been the cutting edge issue on
natural philosophy and science, and in arena of modern physics,
substantial theoretical and experimental efforts are
devoted to unveiling quarks and leptons~\cite{halzen1984}.
Let us focus on the electron that can be isolated stably.
As it stands, one has none of observational evidence of the
element-divided substructure; upper limit of the radius is reported
$10^{-20}\,\mathrm{cm}$~\cite{dehmelt1988,gabrielse2006}.
According to a great success of quantum electrodynamics (QED)
for point particles, we can hardly imagine the electron as
a (composite) particle having finite spread, despite its non-zero
form factor for anomaly of magnetic dipole moment.
The issue of inner structure could be related to the hierarchy
problem~\cite{gildener1976,nilles1984}, which still remains unsolved.
The ``desert'' (e.g., \cite{dimopoulos1990}), compatible with the Large
Hadron Collider experiments for a decade~\cite{canepa2019,aad2021},
implies that any responsive feature might not exist down to the
spatial scale of around $l_{\mathrm{gut}}\sim 10^{-29}\,\mathrm{cm}$
(corresponding to the grand unification energy; e.g., section~93 in
\cite{workman2022}).
From the results of ultra-high-energy cosmic ray experiments
negative for the ``top-down'' models (e.g., \cite{letessier2011}),
there is even a likelihood that the ultramicroscopic featureless
landscape continues down to a scale smaller than $l_{\mathrm{gut}}$.
In short, the Planck length $l_{\mathrm{P}}\sim 10^{-33}\,\mathrm{cm}$
comes into sight.
Unless the electron is made out of any ingredient perceivable in our
space-time, we will have, logically, no choice but to conceive that
the electric charge $-e$ and spin $1/2$ as the observable
attributes reflect structure of another space expanded somewhere
in a virtually infinitesimal region.
The similar spatio-requirement can also be seen in the string
theory that hypothesizes an ingredient having the size of
$\sim l_{\mathrm{P}}$; that is, the theory requires extra-dimensions
of space (e.g., \cite{polchinski1998}).
From the observational fact that magnetic monopole has not been
discovered at all~\cite{eidelman2004,aad2020}, it is simply anticipated
that, the mechanism that generates the magnetic moment coupled
with the intrinsic angular momentum $\hbar/2$~\cite{gerlach1922}, where
$\hbar=h/2\pi$ and $h$ is the Planck constant~\cite{planck1901},
has an asymmetric relation with the mechanism that generates $e$,
while having a complementary relation with it.

For the challenge of elucidating the structure that engenders those
attributes, one of the worthwhile first steps is to find the analogy in
structure of nucleon.
The nucleon is constantly emitting and absorbing $\pi$-meson, to have
the boson cloud (e.g., \cite{friedrich2003}); this picture is similar to
the one of electron in QED in which virtual photon dresses it.
As for orbital angular momentum $\bf L$, the pion cloud entails the
eigenvalue of $l=1$, which means that the pion is
{\em really rotating on an orbit}.
Its connection with the spin angular momentum of electron is
founded on the local isomorphism between $\mathrm{SO(3)}$
and $\mathrm{SU(2)}$ Lie groups; there is no longer any doubt
as to validity of imposing the commutation relation
equivalent to that for the operator of ${\bf L}={\bf R}\times{\bf P}$,
on $\bf S$~\cite{pauli1927}, where the notations are standard.
Accordingly, it will be a decent attempt to envisage, for
generation principle of $\bf S$, a rotational coordinate of the space
expanded in the infinitesimal region of $\|{\bf R}\|\to 0$.
However, primitive question of like what coordinate must be rotated
has hitherto remained unanswered (e.g., \cite{bohm1951}).

It is not totally absurd to look for, further in macroscopic objects,
a clue to the puzzle, since the asymmetry of the electromagnetic
attributes is being cast to predominance of magnetic fields
over cosmological scales (e.g., \cite{durrer2013}).
It is plasma physics that accounts for the dynamics of
many-body system of charged particles coupled with
abelian and non-abelian gauge fields~\cite{tajima1997}.
In the context, it is better to seek out, in magnetohydrodynamical
features of abelian plasma, an intuitive image for the $\bf S$ generation.
Now, we closer look at the astrophysical jets, which are launched from
active galactic nuclei including black holes, to extend splendidly
up to million light years.
Intriguingly, in the jets one can find some signatures of helical
motion as well as helical structure of magnetic field~\cite{broderick2009}.
Such structure could appear when the plasma attains an energy relaxed
state, mostly conserving magnetic helicity~\cite{woltjer1958,taylor1986}.
Actually, helical magnetic field structure observed in, for example,
interplanetary magnetic clouds~\cite{burlaga1988} and plasma ejecta in
the solar corona~\cite{plunkett2000} has been interpreted as
exhibiting the relaxed state.
Therefore, the structure formation is supposed to be a class of
universal magnetohydrodynamic phenomenon.
In this aspect, reminding that the pion can be regarded as the lowest
energy excitation state of vacuum~\cite{nambu1960,goldstone1961},
we genuinely conjecture that the helical structure will reflect the
geometrical structure that generates an internal coordinate
connecting to $\bf S$.
Here, it is amusing to compare the galactic nucleus to one neutron,
invoking the nuclear reaction of $n^{0}\longrightarrow p^{+}+\pi^{-}$,
where $p^{+}$ and $\pi^{-}$ are proton and $\pi^{-}$-meson, respectively:
making a spinning black hole~\cite{akiyama2019} and
accretion disk~\cite{ford1994} correspond, respectively, to $p^{+}$ and its
$\pi^{-}$ cloud leads to the apparent correspondence between a bipolar jet
and lepton pair emitted in decay of $\pi^{-}$, in parallel with the conjecture.
Either way, it seems as if the structure in the infinitesimal space and
outer space was described by common geometry\,---\,this is
original motivation of the present study.

The crucial thing, that divergence difficulty incidental to electron
can be successfully removed by the well-defined renormalization of QED,
implies latency of a fundamental operation allied with transition to the
infinitesimal space.
If Nature is essentially inventive, geometry of the space would be to
provide the physical meaning of degree of charge renormalization,
which has been unknown~\cite{feynman1961}.
Meanwhile, polarization picture of QED vacuum, itself, suggests that
one could by no means reach generation principle of $e$
within theoretical framework tacitly involving the
vacuum permittivity {\it ab ovo}.
This facet is prominent in the logical base of Lorentz invariant
quantum theory of gauge fields~\cite{yang1954} such that
existence of photon with the speed $c$ results necessarily from charge
conservation and local gauge invariance consistent with causality.
Put another way, the inherent problem is that the theory is dependent
on background.
By taking a hint from this impasse {\it per s\={e}}, a na\"{i}ve attempt
is made here to find out a form of the concerned operation: we anticipate
a symptom of that form in the macroscopic dielectric distribution
by which phase velocity of light spatially changes.
The small deviation from $c$ is suggestive of the vacuum polarization
effect.
It is, by contrast with the magnetohydrodynamical analogy for spin
generation, a matter of having an insight into collective phenomena
of plasma as dielectric medium~\cite{anderson1963}, among others,
dispersion of light propagating through the plasma.

Of course, it is impossible, from plasma physics with $\mathrm{U(1)}$
symmetry, to {\em literally} derive the microscopic mechanics
assigning the point the intrinsic quantized attributes.
The mechanics is thought of as being rather inaccessible by
traditional group theories.
For example, the Poincar{\'e} group for relativity provides
classification of elementary particles~\cite{wigner1939},
but fails in determining their mass as an invariant of the
Casimir operator of the group; it all comes down to the
problem of free-particle self-energy.
In this aspect, worth trying is, away from algebraic approach,
an analytic one capable of yielding specific numerical outputs
such as the root of special functions~\cite{feynman1961}.
In the current context, when analytical description of plasmas as
the dielectric and magnetofluid is {\em appropriately generalized and
abstracted}, we may get a fortuitous chance to encounter a hopeful
candidate for the geometry that engenders $e$ and $h$.
If the framework provides radical representation of space-time
independently of background, it would be to encompass
generation mechanism of the dimensionless constant including $c$:
$e^{2}/2hc=e^{2}/q_{\mathrm P}^{2}=\alpha$~\cite{sommerfeld1916},
and its relation to the other interactions.
Mechanics in the infinitesimal space should describe observable
phenomena of particles and fields in the infinite limit.
Thus, priority task is to specify the space-transformation
operation responsible for the reproducibility.

In this {\small Perspective Paper}, we study, for heuristic, the
geometric structure of point(-like, at present stage of knowledge)
particles including the electron, and fundamental interactions
inseparable from them.
Along with the analogy of optical dispersion of plasma,
I propose a simple form of the transformation that describes transition
to the space expanded in the infinitesimal region.
I provide the realistic space, which is spanned by a unific
rotational field equivalent to Gromeka-Beltrami vector
flow~\cite{beltrami1889} with cylindrical geometry.
The infinitesimal realistic space differs from an infinitesimal
element of four-dimensional space-time, and also from
conventional abstract (internal) space such as isospin
space.\footnote[1]{However, I use the terminology ``internal space''
(or ``internal ...'') throughout, for convenience.}
The field is found to obey the rotational eigenvalue equation,
which captures the complex orthogonality between the internal
coordinate space and angular momentum space.
I address that complementary rotation of coordinates in these isotopic
spaces can be reflected in observed spin-precession of electron.
We also argue that the electromagnetic coupling is internally determined
by a left-handed rotational eigenmode, and this property is linked to
parity violation in the $\beta$-decay~\cite{lee1956,wu1957}.
This eigenmode regulates helical structure of the rotational field.
The related geometry is found to be the same as that to describe the
helical structure of the relaxed plasma.
It is pointed out that the geometrical theory is distinct from the
previous morphology in~\cite{bostick1985}, which was constructed
within, basically, the conventional framework of plasma physics.

This paper is organized as follows:
Section~\ref{sec:2} is prepared to give operational tool as a part of
the geometric theory.
We review the optical dispersion of plasma (Sect.~\ref{sec:2.1}),
in order to arrange a rule of the wavenumber transformation of light
accompanying plasma nonuniformity (Sect.~\ref{sec:2.2}).
Taking the abstracted form into account, I propose the spatial
transformation to the infinitesimal space of particles (Sect.~\ref{sec:2.3}).
To see its availability, in~Sect.~\ref{sec:3} we call quark condensate,
as it resembles a massive photon state owing to collective oscillatory
behavior of plasma electrons.
We reproduce the potential detected inside \cite{eichten1978}
(Sect.~\ref{sec:3.1}) and outside meson~\cite{yukawa1935}
(Sect.~\ref{sec:3.2}), clarifying a signature of the infinitesimal
realistic space spanned by harmonic field.
In~Sect.~\ref{sec:4}, I provide an internal module representation of
the $\beta$-decay products.
Making reference to the magnetohydrodynamic details, we geometrically
extend the scalar form, and obtain the rotational eigenvalue equation that
governs vector field spanning that space (Sect.~\ref{sec:4.1}).
The equation is solved as a boundary value problem; we find the
eigenfunctions the lepton pairs should internally refer to
(Sect.~\ref{sec:4.2}).
The eigenmode property including chiral asymmetry is summarized,
and the module representation made of the rotational vector fields
is proposed (Sect.~\ref{sec:4.3}).
The eigenmode representing the weak is also provided (Sect.~\ref{sec:4.4}).
In~Sect.~\ref{sec:5}, we treat major issues on observation of the
rotational field of charged leptons, in terms of application of the
operational formalism to it.
In particular, we focus on geometrical mechanics for cyclotron motion
of a single electron (Sect.~\ref{sec:5.1}).
From the rotational eigenvalue equation governing the field, we derive
a mathematical symbol that sustains the mechanics (Sect.~\ref{sec:5.2}).
Then, we relate the mechanical process renormalizing an internal
coordinate to the charge renormalization (Sect.~\ref{sec:5.3}).
For the sake of reinforcing the theory, Sect.~\ref{sec:6} is added
wherein its compatibility with the standard model is further examined.
First, I provide an explanation of how the mechanics could respond to
fractional charge of quark~\cite{gellmann1964} (Sect.~\ref{sec:6.1}).
Second, we argue the compatibility with the established model of
vacuum~\cite{higgs1964} (Sect.~\ref{sec:6.2}), and third, provide
a notion of rotational eigenmode coupling responsible for the
electroweak coupling~\cite{glashow1961,weinberg1967,salam1968}
(Sect.~\ref{sec:6.3}).
Finally, Sect.~\ref{sec:7} is devoted to concluding remarks.\\

\section{Abelian Plasma Analogy for an Abstract Form of
Spatial Transformation}\label{sec:2}

It is known that classical field theory appropriately provides
physical elements to understand the quantum field theory
(e.g., \cite{sakurai1967}) that any modern approach to theories of
fundamental interactions rests on.
In what follows, propagation of light in spatially nonuniform plasma is
considered, as an analogy for the non-trivial transformation involved in
the novel theory compatible with the quantum field theory.

\subsection{Brief Review of Massive Photon Picture for Plasma Dielectric
Response}\label{sec:2.1}

Begin with considering a discharged gas distributing in vacuum,
which comprises freely moving electrons and the charge compensating ions.
Light propagating through it induces currents carried mainly by the electrons
having the small inertia, whereupon this effect is fed back to the light.
The interaction gives rise to optical dispersion.
For the phenomenological description, one calls the Maxwell equation
in vacuum, taking the free currents into account.
For simplicity, provided that the background ions are immovable,
the current formation is considered which involves harmonic
oscillation of the nonrelativistic electrons that experience a single
force $-e{\bf E}$, where $\bf E$ is the self-consistent electric field.
In this model, one has the Klein-Gordon equation for transverse
electromagnetic field of
$\Phi=\{ {\bf E},{\bf B}\}$~\cite{thomson1927}:
\begin{equation}
\left(\partial^{2}/\partial t^{2}-c^{2}\Delta\right)\Phi
=-(ne^{2}/m_{e})\Phi,
\label{eq:2.1}
\end{equation}
\noindent
where $\Delta=\nabla^{2}$, and $n$ and $m_{e}$ are the plasma density
and the electron rest mass, respectively.
When assuming the plasma and plane wave to be infinitely pervading
the vacuum, equation~\eqref{eq:2.1} immediately gives the dispersion
relation of the electromagnetic wave having its angular frequency
$\omega$ and wavenumber ${\bf k}$:
\begin{equation}
\omega^{2}=c^{2}{\bf k}^{2}+\omega_{p}^{2},
\label{eq:2.2}
\end{equation}
\noindent
where $\omega_{p}$ is the plasma frequency that satisfies
$\omega_{p}^{2}=ne^{2}/m_{e}$~\cite{tonks1929}.
For \eqref{eq:2.2}, the refractive index is given by
$ck/\omega=\sqrt{1-(\omega_{p}^{2}/\omega^{2})}$,
where $k=\|{\bf k}\|$.
On both sides of \eqref{eq:2.2}, we multiply $\hbar^{2}$,
and put $E=\hbar\omega$ and ${\bf p}=\hbar{\bf k}$.
Then, the relativistic relation comes out, to give
$E^{2}=c^{2}{\bf p}^{2}+m^{2}c^{4}$~\cite{einstein1905}, where $m$ is
defined by $m=\hbar\omega_{p}/c^{2}$, which has dimension of mass.
This suggests that the photons behave as though they had the mass,
as in superconductors~\cite{nambu1960,anderson1963}.

The real position vector $\bf R$ of the harmonic oscillator,
which assigns the photons $m$, is parallel to $\bf E$ (perpendicular to
magnetic field $\bf B$), on account of ${\bf R}=(e/k_{s}){\bf E}$, where
$k_{s}=m_{e}\omega^{2}$ can be identified with the spring constant.
This means that, when regarding $k_{s}/e$ as a constant $\Lambda$,
equation~\eqref{eq:2.1} with the following replacement results in describing
the coordinate wave that traces position of the electrons:
\begin{equation}
{\bf E}\,\longrightarrow\,\Lambda{\bf R},\mspace{50mu}
{\bf B}\,\longrightarrow\, i\Lambda{\bf R},
\label{eq:2.3}
\end{equation}
\noindent
where $i=\sqrt{-1}$.
Equation~\eqref{eq:2.3} seems to expose a hidden transformation form
generating internal rotational coordinate of electron, in accord with
the electron picture mentioned above.
Besides, a view of the coordinate wave is reminiscent of
general coordinate transformation of space-time~\cite{einstein1916}
(see~\cite{dirac1975}, for the terminology).

\subsection{Wavenumber Transformation of Light}\label{sec:2.2}

\subsubsection{Elementary Excitation Representation of
a Propagative Massive Photon}\label{sec:2.2.1}

\begin{figure}
\centering
\centerline{\includegraphics[width=11cm]{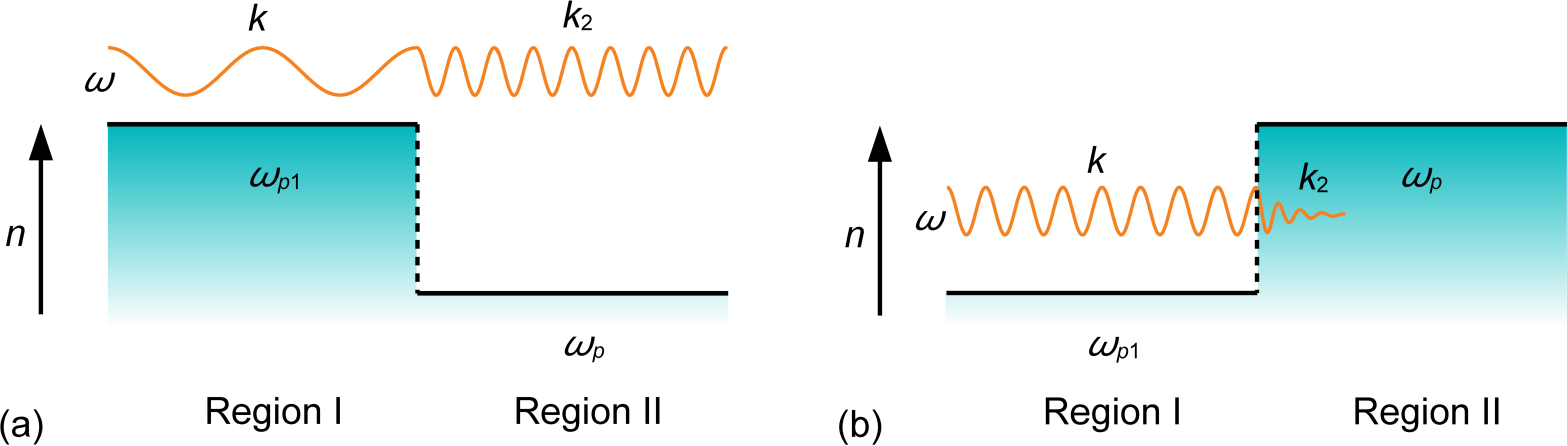}}%
\vspace{.2cm}
\caption{\label{fig:1}Schematics of the wavenumber transformation
$k\to k_{2}$, which comes about when light (wavy solid curves)
propagates from plasma region~I to II across the boundary (dashed lines):
a surface of discontinuity of the density $n$ (solid lines).
Provided the angular frequency of the light $\omega$ is invariant for the
transformation, the cases of $\omega_{p}<\omega_{p1}<\omega$ (a) and
$\omega_{p1}<\omega<\omega_{p}$ (b) are shown, where $\omega_{p1}$
and $\omega_{p}$ denote the plasma frequency of the region~I and II,
respectively.}
\end{figure}

The nonuniformity of plasma, which varies phase velocity of the light,
is taken into consideration.
For instance, as shown in~Fig.~\ref{fig:1}(a), we set two distinct
regions: high density region I specified by the plasma frequency of
$\omega_{p1}$ and low density region II by that of
$\omega_{p}\,(<\omega_{p1})$, in the way that on an infinite
boundary surface, these come in contact with one another.
Monochromatic light with $\omega\,(>\omega_{p1})$ propagating
through the region~I in the direction normal to the contact surface
is transmitted to the region~II across the surface.
Provided that $\omega$ is unchanged in the entire process,
the dispersion relations in the region~I and II can be written as
$\omega^{2}=c^{2}k^{2}+\omega_{p1}^{2}$ and
$\omega^{2}=c^{2}k_{2}^{2}+\omega_{p}^{2}$, respectively.
By combining them, we obtain, for positive real wavenumber
transformation of $k\to k_{2}$, the factorized representation of $k_{2}$:
\begin{equation}
k_{2}=\sqrt{ \left(k^{2}+\frac{\omega_{p1}^{2}}{c^{2}} \right)
\left(1-\frac{\omega_{p}^{2}}{\omega^{2}}\right) }.
\label{eq:2.4}
\end{equation}
\noindent
Note that the second factor of the right-hand side (RHS) indicates
the refractive index: $\tilde{n}=ck_{2}/\omega$ in the region~II.

Here, suppose the situation in which keeping the ratio of
$\omega_{p}/\omega$ constant, $\omega\downarrow\omega_{p1}$
is taken to give $k\to 0$.
For given definitions of $\mu=k_{2}(k=0)$, $\bar{\mu}=\omega_{p1}/c$,
and $\delta=\omega_{p}^{2}/\omega_{\omega_{p1}}^{2}$,
the wavenumber transformation can be expressed as
\begin{equation}
k\left(=0\right)\,\longrightarrow\,\mu=\bar{\mu}\sqrt{1-\delta}.
\label{eq:2.5}
\end{equation}
\noindent
Equation~\eqref{eq:2.5} can be regarded as representing creation of
a propagative photon having the mass of $m=\hbar\bar{\mu}/c$.
Particularly, in the case for which $\delta\ll 1$, the phase velocity
in the region~II, $c/\tilde{n}$, is approximately given by
$c\left(1+\delta/2\right)$, so that it slightly deviates from $c$.
The form is that as we have desired; equation~\eqref{eq:2.5} is
expected to be related with the lowest energy excitation of
real particles.

It is noticed that, when comparing the massive photon to a real matter
particle, we should take the group velocity of the light into consideration.
The lowest energy excitation with
$c^{-1}\|\partial\omega/\partial k\|=\beta\,(<1)\to 0$ can be translated
into unambiguous separation of the kinetic energy
$[(\Gamma-1)mc^{2}]_{\Gamma\to 1}$ from the total energy
$E=[\Gamma mc^{2}]_{\Gamma\to 1}$, where $\Gamma=(1-\beta^{2})^{-1/2}$.
This can be compared to transition from
$\sqrt{E^{2}(\|{\bf p}\|=0)/c^{2}}=mc$ to the nonrelativistic momentum
$[\beta mc]_{\beta\to 0}$.
The formalism suggests that
$\bar{\mu}=\sqrt{\omega^{2}(k=0)/c^{2}}=mc/\hbar$ should be replaced
by $[\beta mc/\hbar]_{\beta\to 0}$ with the value remaining finite.

\subsubsection{Elementary Excitation Representation of
a Non-propagative Massive Photon}\label{sec:2.2.2}

The density profile is inverted with respect to the propagation direction
of the light.
As shown in Fig.~\ref{fig:1}(b), we configure low density region~I
with the plasma frequency of $\omega_{p1}$ and high density region~II
with $\omega_{p}\,(>\omega_{p1})$ such that an infinite contact surface
separates them.
The light with $\omega\,(>\omega_{p1})$ propagating through
the region~I normally incidents on the surface, to be transmitted to
the region~II.
Here, we consider the situation in which the transmitted light decays
because of $\omega<\omega_{p}$.
For the dispersion relations same as given above, we obtain,
for $k\to k_{2}$, the following representation of $k_{2}$:
\begin{equation}
k_{2}=-i\left(\omega_{p}/c\right)\sqrt{1-\left(\omega^{2}/\omega_{p}^{2}\right)},
\label{eq:2.6}
\end{equation}
\noindent
making the negative sign have the physical meaning that a
damping solution has been chosen.
Again keeping the ratio of $\omega/\omega_{p}$,
$\omega\downarrow\omega_{p1}$ is taken to give $k\to 0$,
and then, the wavenumber transformation can be expressed as
\begin{equation}
k\left(=0\right)\,\longrightarrow\,\mu=-i\bar{\mu}^{\ast}\sqrt{1-\delta}
\eqqcolon -i\mu^{\ast},
\label{eq:2.7}
\end{equation}
\noindent
where $\bar{\mu}^{\ast}=\omega_{p}/c$,
$\delta=\omega_{p1}^{2}/\omega_{p}^{2}$, and
$m^{\ast}=\hbar\bar{\mu}^{\ast}/c$, which stands for mass of
non-propagative photon.
Equation~\eqref{eq:2.7} is expected to be related with the
lowest energy excitation of virtual particles.

\subsection{Transformation Between Our Conventional Space and
Infinitesimal Space}\label{sec:2.3}

The wavenumber transformation of the massive photon is coupled
with coordinate of the propagation direction, to generate phase
transformation of $\Phi$ between the region~I and II.
This feature provides a useful analogy for the transformation
between the concerned internal space and the ${\bf R}^{3}$
real-space from which observer looks into the internal one.
The spatial transformation should be related to the generation
of wavefunction of particle, which coincides with potential
development in the ${\bf R}^{3}$.
As seen later, it does {\em not} mean that the transformation
(compared to the scalar form) by itself constitutes
geometrical main body.
Anyway, let us consider a generic form of three-dimensional
isotropic potential $V(\xi^{\prime})$, and suppose that it is
centered at coordinate origin $O: (X,Y,Z)=(0,0,0)$ in the flat
(locally Euclidean) ${\bf R}^{3}$, where the dimensionless,
positive real argument is defined as
$\xi^{\prime}=\bar{\mu}\|{\bf R}\|=\bar{\mu}R\notin\{0,+\infty\}$.
For an equipotential surface of
$R=\sqrt{X^{2}+Y^{2}+Z^{2}}=\xi^{\prime}/\bar{\mu}$, a circle having
the dimensional radius $R$ can be put on the equatorial plane $Z=0$.
This locus is described by the simple blowup:
\begin{equation}
\tilde{X}=\bar{\mu}R\cos\theta^{\prime}
=\xi^{\prime}\cos\theta^{\prime},\qquad
\tilde{Y}=\bar{\mu}R\sin\theta^{\prime}
=\xi^{\prime}\sin\theta^{\prime},
\label{eq:2.8}
\end{equation}
\noindent
and dimensionless equation of the circle is given by
\begin{equation}
\tilde{X}^{2}+\tilde{Y}^{2}={\xi^{\prime}}^{2}.
\label{eq:2.9}
\end{equation}
\noindent
Domain of definition for $V(R)$ is given by $R>\epsilon$: in the domain,
$\bar{\mu}R$ indicates a well-defined real value of $\xi^{\prime}$.
On the other hand, in the region of $R\leq\epsilon$ off the domain,
we suppose that indetermination led by $R\to 0$ sets in,
and require discontinuous transition from \eqref{eq:2.9} to
\begin{equation}
\tilde{X}^{2}+\tilde{Y}^{2}=0,
\label{eq:2.10}
\end{equation}
\noindent
for $\|\tilde{X}\|\nrightarrow 0$ and $\|\tilde{Y}\|\nrightarrow 0$,
as would be a decent representation of null-radius associated
with point particles.
It is trivial that the singular region \eqref{eq:2.10} does not
belong to the Euclidean ${\bf R}^{3}$-space anymore;
the Pythagorean theorem is violated.
Definition of neighbourhood in standard analysis leads to
$\{{\bf R}\mid R<\epsilon\}=\emptyset$, which means that
the Euclidean space is locally punctured.
For the topological modeling, there is a physical reasoning
that could convince us of its feasibility.

We reconsider the non-propagative photon characterized by the real mass
$m^{\ast}(\neq 0)$, real energy $E\,(<m^{\ast}c^{2})$,
and imaginary momentum ${\bf p}_{i}$.
For the relativistic dispersion, instead of normal transformation from
the elliptic to hyperbolic (timelike) form, here we should make the one
to null-vector form, that is, the transition from the virtual particle state
having a real $p^{\ast}=\|{\bf p}_{i}\|$ into massless state:
\begin{equation}
\left(E/c\right)^{2}+p^{\ast 2}=\left(m^{\ast}c\right)^{2}
\,\longrightarrow\,\left(E/c\right)^{2}+p^{\ast 2}=0.
\label{eq:2.11}
\end{equation}
\noindent
For, this can naturally be made to correspond to
\eqref{eq:2.9}$\,\to\,$\eqref{eq:2.10}, albeit apparently breaking
the Lorentz symmetry.
The transition of \eqref{eq:2.11} with $E/c\nrightarrow 0$ and
$p^{\ast}\nrightarrow 0$ actually allows for replacement of
$E/c\to{\bf p}$ and $p^{\ast}\to i{\bf p}$,
with $\bf p$ being a real three-dimensional vector.
The massless particle having the momentum $\bf p$ whose magnitude
indicates $E/c$ is definitely an entity observed as photon in vacuum.
Hence, the \eqref{eq:2.11} accommodated by the replacement
is a physically possible process as spatial transfer from the
dispersive medium to the vacuum.

The singular region in the ${\bf R}^{3}$, characterized by
\eqref{eq:2.10}, is defined as indeterminate region.
In correspondence to what observer can never jump on internal
coordinate of the massless photon, namely, the causal disconnection,
the ${\bf R}^{3}$ spanned by $V(R)$ can not exist in the
indeterminate region due to the definition.
Nevertheless, on the analogy of the replacement by which to access to
the eigenstate of the photon, we can figure out the spatial transformation
to access to internal coordinate intrinsic to particles.
That is, for \eqref{eq:2.9}$\,\to\,$\eqref{eq:2.10}, the following
replacement invoking a basis vector $\hat{\bf x}$ could hold:
\begin{equation}
\tilde{X}\sim\cos\theta^{\prime}\,\longrightarrow\,\hat{\bf x},\qquad
\tilde{Y}\sim\sin\theta^{\prime}\,\longrightarrow\,i\hat{\bf x}.
\label{eq:2.12}
\end{equation}
\noindent
When introducing the orthogonal basis vector $\hat{\bf y}$ that
satisfies $\hat{\bf y}=i\hat{\bf x}$, we obtain $\tilde{Y}\to\hat{\bf y}$.
As is, it turns out that equation~\eqref{eq:2.12} prompts leap from
the real $XY$-plane to Wessel-Argand-Gau{\ss} (complex) plane,
settling $\hat{\bf x}$ and $\hat{\bf y}$ as the axes.
This plane exists in ${\bf r}^{3}\times{\bf r}^{3}$ host space as
the {\em point} (as an element) of complex three-dimensional space:
$\hat{\bf x}+i\hat{\bf y}={\bf 0}$, sustaining orthogonality of
the two ${\bf r}^{3}$-real spaces.
Note here that the ${\bf r}^{3}$ should be distinguished from ${\bf R}^{3}$,
as is unable to define the norm $\|{\bf R}\|$.
The internal ${\bf r}^{3}$- or ${\bf r}^{3}\times{\bf r}^{3}$-space
is referred to as infinitesimal realistic space (or infinitesimal space);
the peculiar complex space $\{\bf 0\}$ is supposed to be
in the indeterminate region.
In terms of the specific setting, therefore, the infinitesimal space
is distinguished from conventional compactified space,
though their geometrical connection may be anticipated
(cf. Sect.~\ref{sec:5.3}).

On the complex plane, we have $\zeta=\tilde{x}+i\tilde{y}$ with
$\tilde{x}$ and $\tilde{y}$ being real, and give
$(\tilde{x},\tilde{y})=(\xi\cos\theta,\xi\sin\theta)$
as with \eqref{eq:2.8}.
For $(\tilde{X},\tilde{Y})\to(\tilde{x},\tilde{y})$, we suppose the
linear transformation of
$(\theta^{\prime},\xi^{\prime})\to(\theta,\xi)$ of the form:
$\theta^{\prime}/g_{0}\to\theta$ and $\xi^{\prime}/g_{0}\to\xi$ with
$g_{0}=2$, in consideration of the multiplicity of ${\bf r}^{3}$ and
loop continuity of $2\pi\xi^{\prime}=g_{0}(2\pi\xi)$.
Putting $\xi=kr$ provided $\zeta\in\mathbb{C}_{\ast}
\subseteq\{\tilde{z}\in\mathbb{C}\mid\tilde{z}\ne 0,\infty\}$,
$\xi^{\prime}\to g_{0}\xi$ is cast to the expression of
$\bar{\mu}R(=\infty\cdot 0)\,\longrightarrow\,g_{0}kr$,
where the left-hand side (LHS) including the round bracket represents
that for $R\to 0$ and $\bar{\mu}\to\infty$, $\bar{\mu}R$ is
in the indeterminate region, manifesting extension of the
number system of
$\mathbb{R}_{\ast}\coloneqq\{-\xi^{\prime},\xi^{\prime}\}
\subset\mathbb{R}$.
Noted is that $\mathbb{C}_{\ast}$ stands for an open subset of
the finite complex numbers, and $\mathbb{R}_{\ast}$ the open set
of the finite real numbers.
Formally introducing $g=g_{0}\sqrt{1-\delta}$ with $\delta$
being a parameter, we have
\begin{equation}
\mu R\left(=\infty\cdot 0\right)\,\longrightarrow\,gkr,
\label{eq:2.13}
\end{equation}
\noindent
where $\mu=\bar{\mu}\sqrt{1-\delta}$.
Equations~\eqref{eq:2.12} and \eqref{eq:2.13} can be understood as a
feasible form of transformation from the ${\bf R}^{3}$ space to the
infinitesimal ${\bf r}^{3}\times{\bf r}^{3}$ space, along the
extension: $\mathbb{R}_{\ast}\cup\{\pm\infty\}
\left(\supset\bigl\{\frac{\pm\xi^{\prime}}{\pm\infty}\bigr\}\right)
\to\mathbb{C}_{\ast}$
or $\mathbb{R}_{\ast}\cup\{\pm 0\}\to\mathbb{C}_{\ast}$,
to be involved in transformation from potential function
$V(\xi^{\prime})$ to an internal function dependent on $\xi$.
Here, it is supposed that the extended sets belong to
the Affinely extended real number
$\bar{\mathbb{R}}=[-\infty,+\infty]$, paralleling the
extension of $\mathbb{C}_{\ast}$ given below.

We now consider the inverse transformation:
${\bf r}^{3}\times{\bf r}^{3}\to{\bf R}^{3}$
that generates $V(\xi^{\prime})$ as an output function.
For the inverse of \eqref{eq:2.12}, we read off the following
replacement, invoking a basis vector $\hat{\bf X}$:
\begin{equation}
\tilde{x}\sim\cos\theta\,\longrightarrow\,\hat{\bf X},\qquad
i\tilde{y}\sim i\sin\theta\,\longrightarrow\,-\hat{\bf X}.
\label{eq:2.14}
\end{equation}
\noindent
As the inverse of \eqref{eq:2.13} in which \eqref{eq:2.14} is involved,
a possible form is
\begin{equation}
kr\left(=0\cdot\infty\right)\,\longrightarrow\,g^{-1}\mu R.
\label{eq:2.15}
\end{equation}
\noindent
Here, the LHS including the round bracket represents that for $k\to 0$
and $r\to\infty$, $kr$ is in indeterminate region, manifesting
the extension of $\mathbb{C}_{\ast}$ (in the conventional sense).
Equations~\eqref{eq:2.14} and \eqref{eq:2.15} suggest that in the
infinite distance of ${\bf r}^{3}$ there exists ${\bf R}^{3}$, along
$\mathbb{C}_{\ast}\cup\{0\}\left(\supset\bigl\{\frac{\xi}{0}\bigr\}\right)
\to\mathbb{R}_{\ast}$ or $\mathbb{C}_{\ast}\cup\{\infty\}
\to\mathbb{R}_{\ast}$.
The limit of $k\to 0$, cooperating with $r\to\infty$, can be interpreted
as a projection linked to the foregoing setting of $Z=0$ in ${\bf R}^{3}$.

Equation~\eqref{eq:2.15} appears to be of the form that enables us to
make $k\,(=0)$ and $\mu(\bar{\mu},\delta)$ correspond to those
in \eqref{eq:2.5}, and $r=\infty$ to a bulk condition for
the plane wave to have the diverging wavelength.
As is, we also prepare $\mu(\bar{\mu}^{\ast},\delta)$ with the
factorized form same as \eqref{eq:2.7}, though $\bar{\mu}$,
$\bar{\mu}^{\ast}$, and $\delta$ herein are, at this moment, 
regarded as the abstracted symbols.
Here, we pay attention to what in \eqref{eq:2.5} $k=0$ was needed
for fixing $\bar{\mu}$, whereas in \eqref{eq:2.7} $\bar{\mu}^{\ast}$
was free of $k$.
Besides, since the previous $\delta$ could be handled as parameter
intrinsic to the elementary excitation in the double plasma,
namely plasmon, the corresponding symbolic $\delta$ as well
is supposed to be a parameter intrinsic to elementary excitation
in double (causally disconnected) vacuum.
Taking all this into consideration will now convince us of physical
relevance of the following proposition:
the ${\bf r}^{3}\times{\bf r}^{3}\to{\bf R}^{3}$ transformation related to
the lowest energy excitation of real and virtual particle,
respectively represented in the form of
\begin{subequations}
\label{eq:2.16}
\begin{align}
kr\bigl(=0^{(\prime)}\cdot\infty\bigr)&\longrightarrow
g_{0}^{-1}\bar{\mu}R^{\ast},
\label{eq:2.16a}
\\
kr\bigl(=0\cdot\infty^{(\prime)}\bigr)&\longrightarrow
g_{0}^{-1}\left(-i\bar{\mu}^{\ast}\right)R^{\ast}=g_{0}^{-1}\bar{\mu}^{\ast}R,
\label{eq:2.16b}
\end{align}
\end{subequations}
\noindent
where the primes in \eqref{eq:2.16a} and \eqref{eq:2.16b} signify that
$k\to 0$ and $r\to\infty$ take the lead in the indetermination,
along $\mathbb{C}_{\ast}\cup\{0\}$ and
$\mathbb{C}_{\ast}\cup\{\infty\}$, respectively.
The quantities $\bar{\mu}$ and $\bar{\mu}^{\ast}$ are related to
observable mass $m$ of real and virtual particle, by
$\bar{\mu}=[\beta mc/\hbar]_{\beta\to 0}$ and
$\bar{\mu}^{\ast}=mc/\hbar$, respectively, where
$\beta=c^{-1}(\partial E/\partial p)$.

In general, distinguishing indeterminate forms (and comprehensive
treatment of number systems) could be a subject of
nonstandard analysis that invokes hyperreal number.
Concerning this, an intuitive explanation for the LHS of \eqref{eq:2.16}
is provided by employing, e.g., the Bessel function
$J_{1}(\xi)$:\footnote[2]{Notation of special functions
is as with that in \cite{watson1958} throughout.}
Adding the points $\{0\}$ and $\{\infty\}$ that never accommodate
$J_{1}(\xi=0)=0=J_{1}(\xi=\infty)$ because of $\xi\notin\{0,\infty\}$,
we allow for $J_{1}(k\to 0)\sim\xi/2$ and $J_{1}(r\to\infty)\sim
\sqrt{2/\pi\xi}\cos(\xi-3\pi/4)$, which correspond to the regimes of
$\eqref{eq:2.16a}$ and $\eqref{eq:2.16b}$, respectively.
In this sense, the indetermination is apparent in the internal space,
and standard analysis is still legitimate.
Relating to this, the foregoing indeterminate form
for $\bar{\mu}R$ may be written as
$\bar{\mu}R\bigl(=\infty\cdot 0^{(\prime)}\bigr)$,
in conformity with \eqref{eq:2.16}.
This realm would postulate the smaller mass of
$\bar{\mu}\ll l_{\mathrm{P}}^{-1}$, viz.,
$R_{\mathrm{S}}\ll\sim l_{\mathrm{P}}$, where $R_{\mathrm{S}}$
is the Schwarzschild radius~\cite{schwarzschild1916}.
Another realm of $\bar{\mu}R\bigl(=\infty^{(\prime)}\cdot 0\bigr)$,
which implies larger mass, is not discussed here.\\

\section{Feasibility of the Transformation and a Signature of the
Infinitesimal Space}\label{sec:3}

\subsection{Complex Analytical Consideration on
Quark-Antiquark Potential}\label{sec:3.1}

Prior to application of the spatial transformation to a single fermion,
we try exploiting it for a quark pair as a Bose-Einstein condensate
closer with the plasmon mentioned above.
This orientation is taken to clarify technical feasibility of the
transformation operation and, at the same time,
a signature of the infinitesimal realistic space.
While the composite particle has a finite size as large as classical
electron radius, it just embodies a closed ${\bf R}^{3}$ real-space
perfectly shielded from external (our macroscopic) ${\bf R}^{3}$-space.
The feature confirmed experimentally is found to be well captured by
the transformation~\eqref{eq:2.16}.
For convenience, we begin with recalling the $\beta^{-}$-decay reaction:
$\nu_{e}+d\longrightarrow e^{-}+u$, where the notations are standard.
For including $\beta^{+}$-decay, denoting by $q$ ($\bar{q}$) quark
(antiquark) and by $\ell$ ($\bar{\ell}$) lepton (antilepton),
we here give the generic expression of
\begin{equation}
q_{\mathrm L}+(\bar{q})_{\mathrm R}\,\longrightarrow\,
\ell_{\mathrm L}+(\bar{\ell})_{\mathrm R},
\label{eq:3.1}
\end{equation}
\noindent
where the subscripts $\mathrm L$ and $\mathrm R$ stand for
the left- and right-handed state, respectively.

We aim at associating $k\to 0$ followed by \eqref{eq:2.16},
with the lowest energy excitation of quarks.
For the moment, this is intended for a $q\bar{q}$ pair whose motion
is nonrelativistic so that $\beta\to 0$, though the approximation
is known to be practically better for heavier mesons, typically quarkonia.
We call a basic form of the $q\bar{q}$ potential that has been confirmed
for such mesons, namely, the Cornell potential~\cite{eichten1978}
\begin{equation}
V(R^{\ast})=C_{1}R^{\ast}-C_{-1}R^{\ast -1},
\label{eq:3.2}
\end{equation}
\noindent
where $C_{1}$ and $C_{-1}$ are real constants, and
$R^{\ast}=\|{\bf R}^{\ast}\|$ is distance between $q$ and $\bar{q}$
in the closed space confining colors.
An attention is paid to the fact that this function
satisfies the ordinary differential equation of
\begin{equation}
\left(d^{2}/dR^{\ast 2}+R^{\ast -1}d/dR^{\ast}-R^{\ast -2}\right)V=0.
\label{eq:3.3}
\end{equation}
\noindent
Let us see the bracket of \eqref{eq:3.3} as an operator acting on $V$.
Then, it seems as if exposing a part of the Laplacian for cylindrical
coordinate system.
In fact, the function of \eqref{eq:3.2} with $R$ replacing $R^{\ast}$
amounts to radial part of solution of the Laplace equation in ${\bf R}^{2}$
cylindrical system $(R,\Theta)$ uniform in axial $Z$-direction,
suggesting a specific choice of axes in the ${\bf R}^{2}$ space
such as $\cos\Theta=1$.
On the other hand, we will see no convincing reason for regarding $R$
as radial coordinate of the fictitious cylinder furnished in our
${\bf R}^{3}$ space.
At this juncture, we virtually introduce ${\bf r}^{3}$ real space
to configure the cylindrical system.
Thereupon, we presume the harmonic function $\phi({\bf r})$
in the space, which satisfies $\Delta\phi(r,\theta,z)=0$.
Concerning the solution of the form of $\phi=\eta(r)e^{i(m\theta-kz)}$,
we contemplate the equation for $\eta$:
\begin{equation}
\left(d^{2}/dr^{2}+r^{-1}d/dr-m^{2}/r^{2}-k^{2}\right)\eta=0.
\label{eq:3.4}
\end{equation}
\noindent
For the bracket set to $m=\pm 1$, taking $k\to 0$ compared to
$\beta\to 0$, and simultaneously carrying out replacement of
$r\to R^{\ast}$, bring about the operator of \eqref{eq:3.3}.
Evidently, this procedure entails the function transformation
of $\eta(r)\to V(R^{\ast})$.

This is not mathematical artifact.
Real part of the complex function of $\phi(k\to 0)$, i.e.,
$\eta\cos\theta$ for $m=\pm 1$, can be virtually compared to
electrostatic potential in the ${\bf R}^{2}$ cylindrical free-space,
especially, that proportional to $R\cos\Theta$;
$\|\phi\|=\eta$ corresponds to the potential amplitude.
On the complex plane spanned by the scalar field $\phi(k\to 0)$,
one can choose a projective axis along which to put the basis vector
$\hat{\bf X}$, such that $\cos\theta\to\hat{\bf X}$.
Then, $\eta\hat{\bf X}$ analogous to displacement of circular membrane
oscillation (cf. Sect.~\ref{sec:4.1}) could be regarded as a
coordinate vector (like a generalized coordinate in Lagrangian,
associated with scalar fields; e.g., \cite{sakurai1967}).
Meanwhile, color electric potential binding $q\bar{q}$ is approximately
given by $V(R^{\ast})\propto R^{\ast}$ for large
$R^{\ast}$~\cite{chew1962}.
It is thus found that transformation of the potential amplitude,
$\|\phi\|\to V$, can be related to generation of the rotational coordinate
along $\eta\hat{\bf X}\to{\bf R}^{\ast}\!/2$.
As for quarks of pion, quadratic (harmonic oscillatory) feature of a
rotational parameter of chiral transformation appears in
isospin-space~\cite{gellmann1968}.
In the context, we envisage that the potential established in the
${\bf R}^{3}$ space reflects $\eta^{2}$ norm;
the relevance is seen later.
It should be, here, noted that $k\ne 0$ is necessary for ensuring
${\bf r}^{3}\ne{\bf R}^{3}$, whereas $k\to 0$ with
$\hat{\bf z}\coloneqq{\bf k}/k\to{\bf 0}/0$ can be compared to the
uniformity in an indeterminate $Z$-direction.

In the \eqref{eq:3.4} divided by $k^{2}$, we put $\xi=kr\notin\{0,\infty\}$.
In the general case for which $m$ is integer, two linearly independent
solutions to this equation are the modified Bessel functions of order
$m$, $I_{m}(\xi)$ and $K_{m}(\xi)$ [in the case of $m\neq$ integer,
$I_{m}(\xi)$ and $I_{-m}(\xi)$].
Hence, $I_{\pm 1}(\xi)$ and $K_{\pm 1}(\xi)$ are in on the current issue.
Their asymptotic forms (leading contributions) for the small $\xi$,
in which $k\to 0$ takes the lead, should be considered each;
this secures validity of the linear combination form over the
concerned entire $\xi$-range.
It then follows that $k\to 0$ signifies the projection onto cylindrical base.
The scalar function $\phi(k\to 0)$ as the linear combination of the
asymptotic forms constitutes a complex function, which is denoted by
$f(\zeta)\eqqcolon w$.
Here, $\zeta=\xi e^{i\theta}$ and
$(\xi\cos\theta,\xi\sin\theta)=(\tilde{x},\tilde{y})$,
along the notation in Sect.~\ref{sec:2.3}.

Letting $u(\xi,\theta)$ and $v(\xi,\theta)$ be real functions,
we assume the complex function of $w=u+iv$ to be analytic.
Then, $u$ and $v$ satisfy the Cauchy-Riemann equation,
\begin{equation}
\partial u/\partial\xi=\xi^{-1}\partial v/\partial\theta,\qquad
\xi^{-1}\partial u/\partial\theta=-\partial v/\partial\xi,
\label{eq:3.5}
\end{equation}
\noindent
to be the harmonic conjugates on the $\tilde{x}\tilde{y}$-plane.
Among the linear combinations, a feasible form is found to be
$c_{1}\lim_{k\to 0}\left[I_{1}(\xi)e^{i\left(\theta-kz\right)}\right]
-c_{-1}\lim_{k\to 0}\left[K_{-1}(\xi)e^{i\left(-\theta-kz\right)}\right]$,
where $c_{1}$ and $c_{-1}$ are real constants.
That is, we have
\begin{equation}
u=\left(\frac{c_{1}}{2}\xi -c_{-1}\frac{1}{\xi}\right)\cos\theta,\quad
v=\left(\frac{c_{1}}{2}\xi +c_{-1}\frac{1}{\xi}\right)\sin\theta,
\label{eq:3.6}
\end{equation}
\noindent
which suffice for \eqref{eq:3.5}.
This indicates that the two modes of $m=\pm 1$ contribute to $w$;
these call for ${\bf r}^{3}\times{\bf r}^{3}$ spaces, to be
intertwined via the complex plane.
In the regime led by $k\to 0$, the variable $\xi$ can be expressed as
$kr\bigl(=0^{(\prime)}\cdot\infty\bigr)$, along \eqref{eq:2.16a}.
Of importance is to clarify the domain of $\xi$ for \eqref{eq:3.6},
that is, the upper limit of $\xi$ by which $r\to\infty$ is well constrained.
The $f(\zeta)$ provides the following form of linear map
from the complex plane $\zeta$ to $w$:
\begin{equation}
w=\left(c_{1}/2\right)\zeta-c_{-1}\zeta^{-1}.
\label{eq:3.7}
\end{equation}
\noindent
Note here that the coordinate transformation
$f\colon(\tilde{x},\tilde{y})\to (u,v)$ exhibits singularity at
$\tilde{k}=ka=\sqrt{2c_{-1}/c_{1}}$.
For the transformation to be bijective conformal map,
$\xi<\tilde{k}$ must be satisfied, so that $\tilde{k}$ can be
regarded as the upper limit of $\xi$.
In particular, circular rotation on $\xi=\tilde{k}$ is, in the case of
$c_{1}c_{-1}>0$, transformed to oscillation on line segment of
imaginary axis connecting $(u,v)=(0,-\sqrt{2c_{1}c_{-1}}\,i)$ to
$(0,\sqrt{2c_{1}c_{-1}}\,i)$ on the $w$-plane.
On the other hand, in the case of $c_{1}c_{-1}<0$ (including the
Kutta-Joukowsky transform~\cite{joukowsky1910}),
circular rotation on $\xi=\sqrt{-2c_{-1}/c_{1}}$ is transformed to
oscillation on line segment of real axis connecting
$(-\sqrt{-2c_{1}c_{-1}},0)$ to $(\sqrt{-2c_{1}c_{-1}},0)$.
As is, two-valuedness arises on the real axis of the $w$-plane.
It turns out that such a finite domain prohibited for real $u$
does not appear in the former case; therein we give
specific definition of $\mathbb{C}_{\ast}$ as
$\{\zeta\mid 0<\xi<\tilde{k}\}$.

In the case for which $c_{1}c_{-1}>0$, the real function of
$u=\Re(w)$ could appear as the potential~\eqref{eq:3.2}.
It is natural to identify the concerned ${\bf r}^{3}$-space with
the one that has been provided in Sect.~\ref{sec:2.3}.
For generation of $V(R)$, the lead by $k\to 0$, along
$\mathbb{C}_{\ast}\cup\{0\}$, should be involved in
the transformation of ${\bf r}^{3}\times{\bf r}^{3}\to{\bf R}^{3}$.
Recalling \eqref{eq:2.14}, $u$ in \eqref{eq:3.6} is rewritten as
$\eta(\xi)\hat{\bf X}$, where $\eta$ stands for real amplitude of
the intertwined modes.
Now, the transformation of which the type is the same as \eqref{eq:2.16a}
\begin{equation}
\xi\coloneqq kr\bigl(=0^{(\prime)}\cdot\infty\bigr)\,
\longrightarrow\,g_{0}^{-1}\bar{\mu}R^{\ast}
\label{eq:3.8}
\end{equation}
\noindent
is applied to $\eta(\xi)$.
The generated function for $R^{\ast}$ is then found to show the form of
\eqref{eq:3.2} the nonrelativistic quantum chromodynamics (QCD) is
accountable to (e.g., \cite{bali2001}).

\subsection{The Yukawa Potential}\label{sec:3.2}

If two-valuedness is prohibited on the $w$-plane, $\xi$ cannot
exceed $\tilde{k}$, as long as $k\to 0$ takes the lead.
This seems to capture characteristic of quark confinement;
the $\zeta$-dependence of $w$ to be reflected in the
energy-scale dependence of the Wilson's law~\cite{wilson1974},
that is, the first (second) term of the RHS of~\eqref{eq:3.7},
in the area (perimeter) law.
In order to move to region of larger $\xi$, we should prepare
$\phi(r\to\infty)$ owing to the lead of $r\to\infty$:
the linear combination of the asymptotic forms of
$c_{1}\lim_{r\to\infty}\left[I_{1}(\xi)e^{i\left(\theta-kz\right)}\right]
-c_{-1}\lim_{r\to\infty}\left[K_{-1}(\xi)e^{i\left(-\theta-kz\right)}\right]$.
That is, we have
\begin{equation}
\frac{c_{1}e^{\xi}-\pi c_{-1}e^{-\xi}}{\sqrt{2\pi\xi}}\cos\theta
+i\frac{c_{1}e^{\xi}+\pi c_{-1}e^{-\xi}}{\sqrt{2\pi\xi}}\sin\theta,
\label{eq:3.9}
\end{equation}
\noindent
with $\xi$ being $kr\bigl(=0\cdot\infty^{(\prime)}\bigr)$, along \eqref{eq:2.16b}.
The indeterminate form manifests that $k\to 0$ is taken to the extent
that the large $\xi$ is well maintained.
In this regime, it is found, especially for $c_{1}/c_{-1}=\pi$, that
equation~\eqref{eq:3.9} can be cast to the next linear combination
form of
\begin{equation}
\left\{
\begin{array}{c}
c_{1}\\
\pi c_{-1}
\end{array}
\right\}
\left[I_{1/2}(\xi)\cos\theta +iI_{-1/2}(\xi)\sin\theta\right].
\label{eq:3.10}
\end{equation}
\noindent
This indicates that the system spanned by $\phi(r\to\infty)$
is accommodated by $\theta\to\theta^{\prime}/g_{0}$.
As is, we again apply \eqref{eq:2.14} to \eqref{eq:3.10},
for $\mathbb{C}_{\ast}\cup\{\infty\}$, where
$\mathbb{C}_{\ast}=\{\zeta\mid\tilde{k}<\xi<\infty\}$.
The resulting quantity is written as $\eta(\xi)\hat{\bf X}$ as before,
and then, the function $\eta$ is given by
\begin{equation}
\eta(\xi)=-2
\left\{
\begin{array}{c}
c_{1}/\pi\\
c_{-1}
\end{array}
\right\}
K_{\pm 1/2}(\xi)\eqqcolon\frac{c_{0}}{\surd\pi}K_{\pm 1/2}(\xi).
\label{eq:3.11}
\end{equation}
\noindent
Along the discussion in the previous section, $\eta^{g_{0}}$ with
$g_{0}=2$ is translated as internal Hamiltonian density of
the harmonic oscillation.
To the norm, we apply the transformation as with \eqref{eq:2.16b}.
This results in yielding the energy transformation of
\begin{equation}
\eta^{g_{0}}(\xi)\,\longrightarrow\,\alpha_{s}
\frac{e^{-\bar{\mu}^{\ast}R}}{\bar{\mu}^{\ast}R}\eqqcolon
\frac{V(R)}{m_{\pi}c^{2}},
\label{eq:3.12}
\end{equation}
\noindent
with $c_{0}^{g_{0}}\to\alpha_{s}$.
Here, $\alpha_{s}$ is related to the dimensionless constant of
$\vg_{s}^{2}/q_{\mathrm P}^{2}$, where $\vg_{s}$ denotes the coupling
strength between pion and nucleon, and $\bar{\mu}^{\ast}=m_{\pi}c/\hbar$
with $m_{\pi}$ being the observed pion mass.

According to asymptotic property of
$K_{\pm 1}(kr)\sim K_{\pm 1/2}(kr)$ for $r\to\infty$, the product
$\sim c_{0}^{2}K_{-1/2}K_{1/2}$ can be understood as far-field
interference between $K_{-1}$ and its counterpart $K_{1}$.
In this connection, it is remarked that the same mathematical
representation appears in an attractive interaction of plasma elements;
formation and interference of dipole magnetic fields characterized by
$K_{\pm 1}$ can be seen in self-organization of vortices as
two-dimensional structure of plasma turbulence~\cite{hasegawa1977}
and merging of the vortices~\cite{honda2000}.
Intermittent activity of galactic nuclei is expected to establish
this kind of turbulent state~\cite{silva2003,honda2009},
likely followed by transition to an energy relaxed state.

The function of $V(R)$ exposes the potential energy for strong nuclear
force~\cite{yukawa1935} being proportional to the scalar field of pion
with spin $0$.
That inertia of the constituent quarks is so small as to degrade
nonrelativistic approximation for the motion seems to be reflecting the
constraint on $k\to 0$, contrast to the previous constraint on $r\to\infty$.
Relating to this, $\xi\gtrsim\mathcal{O}(1)$ can be reflected in the region
of $R\gtrsim\bar{\mu}^{\ast-1}$ in which the Yukawa potential appears.

The results obtained in this section support physical validity of
introducing, in the infinitesimal space, the cylindrical system with the
axis along which to take $k\to 0$, and the related harmonic functions.
The number of $\|m\|=1$ herein can be identified with magnitude
of spin of gluons.
What $k=0$ is prohibited in the system of \eqref{eq:3.4}
suggests that quarks could never be at rest.\\

\section{Rotational Field in the Cylindrical System}\label{sec:4}

\subsection{Scalar Form on the Base, and the Geometric
Extension}\label{sec:4.1}

From the internal potential representation, we should derive
internal representation of (A) onset of the scattering in \eqref{eq:3.1}
and (B) the created lepton pairs.
To achieve the objective, it will be better to take a strategy different
from drawing a representation of {\em baryonic} fermion from the
one of pion~\cite{skyrme1961} (and also, \cite{wess1971,witten1983}),
although the internal pion field $K_{\pm 1/2}$ is the same with far-field
of quantum vortex~\cite{abrikosov1957} as a topological soliton.
In this aspect, is suggestive the energy relaxation and
self-organization of plasma in which topological change in ${\bf R}^{3}$
owing to magnetic reconnection~\cite{parker1957,petschek1964}
is known to play a leading r\^{o}le (e.g., \cite{taylor1986}).
Specifically, the complexity of a column magnetofluid is expected
to be in on our geometrical development.
The detailed numerical studies reveal that the relaxation takes place
typically in two steps~\cite{horiuchi1986,sato1996};
the first establishes a quasi-equilibrium configuration dominated
by a direct current mode (uniform in $Z$-axis), and the second,
the helical configuration as the energy minimum state.
In reference to this, we make a heuristic analogy between
the former (latter) configuration and the internal harmonic field
limited (not limited) to $k\to 0$.
Along with this, we consider the internal topological change
that can be decomposed into two processes: (A$^{\prime}$) the
(complex-)homeomorphic and (B$^{\prime}$) heteromorphic
transformation as the representation of (A) and (B), respectively.

For (A$^{\prime}$), we make a simple attempt at transforming the
elliptic partial differential equation for $\phi$ into a hyperbolic one,
on the base.
Practically, this can be performed by the transformation form of
$k(=0)\to -i\mu^{\ast}$ with $m$ unchanged.
When it is applied to $\eta\sim I_{\pm 1}(kr)$, which must be, at
$r\to\infty$, internally coexisting with the $K_{\pm 1/2}(kr)$,
we have $I_{\pm 1}(kr)\to\mp iJ_{\pm 1}(\mu^{\ast}r)$,
where the double signs correspond; this can be interpreted as an
internal representation of gauge field transformation.
The equation that $\eta(\mu^{\ast}r)$ obeys is then given by
$\left(\mu^{\ast -2}\nabla_{\perp}^{2}+1\right)\eta=0$,
where $\mu^{\ast}\ne 0$.
The function $\eta$ is included in the scalar function
$\phi=\eta(\mu^{\ast}r)e^{i(\pm\theta-kz)}$ as a solution of the
following equation in the cylindrical system:
\begin{equation}
\left(\Delta+k^{2}+\mu^{\ast 2}\right)\phi(r,\theta,z)=0.
\label{eq:4.1}
\end{equation}
\noindent
Here, $k$ is newly introduced of which the value can be taken arbitrarily.
Finite value of $k$ is indicative of onset of the well-defined
$\hat{\bf z}\,(={\bf k}/k)$.
Especially for $k=0$, $\eta$ represents single-valued amplitude
of vibration of a circular membrane on $z$-constant slices.
In the current context, regarding $\eta^{g_{0}}(\mu^{\ast}r)$
as the internal potential that acts on $\ell\bar{\ell}$ pair,
it is appropriate to interpret the $\phi$ as a possible representation
of internal scalar field of the weak boson; $\|m\|=1$ and $\mu^{\ast}$
could be reflected in the spin and mass, respectively
(the extended argument is given later in Sect.~\ref{sec:6}).

In the region that allows for
$\xi\coloneqq \mu^{\ast}r\bigl(=0\cdot\infty^{(\prime)}\bigr)$,
$\eta\sim\mp iJ_{\pm 1}(\xi)$ can be expressed as
$-i[J_{-1/2}(\xi)-J_{1/2}(\xi)]/\surd{2}$, where $\xi\notin\{0,\infty\}$
is redefined in the transformed system of $m=\pm 1/2$.
As concerns coupling of the virtual boson with leptons, we see that
modules including the spherical Bessel functions $J_{\mp 1/2}$ are
generated by applying $k(=0)\to -i\mu$ to the internal pion field.
That is, we have
$K_{\mp 1/2}(kr)\to(i\pi/2)e^{\mp i\pi/4}H_{\mp 1/2}^{(1)}(\mu r)$
for the reaction~\eqref{eq:3.1}, where
\begin{subequations}
\label{eq:4.2}
\begin{align}
&H_{-1/2}^{(1)}e^{-i\theta/2}=\left[J_{-1/2}(\mu r)
+iJ_{1/2}(\mu r)\right]e^{-i\theta/2},
\label{eq:4.2a}
\\
&iH_{1/2}^{(1)}e^{i\theta/2}=\left[J_{-1/2}(\mu r)
+iJ_{1/2}(\mu r)\right]e^{i\theta/2},
\label{eq:4.2b}
\end{align}
\noindent
and the complex conjugates are
\begin{align}
&\left(H_{-1/2}^{(1)}\right)^{\ast}e^{i\theta/2}
=\left[J_{-1/2}(\mu r)
-iJ_{1/2}(\mu r)\right]e^{i\theta/2},
\label{eq:4.2c}
\\
&\left(iH_{1/2}^{(1)}\right)^{\ast}e^{-i\theta/2}
=\left[J_{-1/2}(\mu r)
-iJ_{1/2}(\mu r)\right]e^{-i\theta/2},
\label{eq:4.2d}
\end{align}
\end{subequations}
\noindent
respectively.
Here, $\|m\|=1/2$ is compared to spin of leptons, and all terms should be
multiplied by $e^{-ikz}$ with $k$ newly introduced (as before).
The representation with $k=0$ is compared to rest state of them,
whereas $k\neq 0$ to the moving state.
For the case in which $k\neq 0$, hereafter we set to as $k>0$
without loss of generality.

When interpreting \eqref{eq:4.2a}\,$\rightleftarrows$\,\eqref{eq:4.2c}
and \eqref{eq:4.2b}\,$\rightleftarrows$\,\eqref{eq:4.2d} as the
$CP$ transformation operation, we can associate \eqref{eq:4.2a} and
\eqref{eq:4.2b} with $e^{-}_{\mathrm L}$ and $(\bar{\nu}_{e})_{\mathrm R}$,
respectively, emitted in the $\beta^{-}$-decay, and \eqref{eq:4.2c} and
\eqref{eq:4.2d} with $e^{+}_{\mathrm R}$ and $(\nu_{e})_{\mathrm L}$,
respectively, in the $\beta^{+}$-decay.
An asymptotic form of radially outgoing s-wave function can be
obtained by applying
$\mu r\bigl(=0\cdot\infty^{(\prime)}\bigr)\to g_{0}^{-1}\bar{\mu}R$
to $\bigl(H_{-1/2}^{(1)}\bigr)^{g_{0}}$, that is,
$(2/\pi)e^{2i\mu r}/\mu r\to\sim e^{i\bar{\mu}R}/\bar{\mu}R$,
where $\bar{\mu}=[\beta m_{e}c/\hbar]_{\beta\to 0}$ stands for
the observed wavenumber of nonrelativistic electron.
For internal coupling of matter with gauge field, we require
$\xi(\coloneqq\mu^{\ast}r)=\mu r$ on the cylindrical base.
Here, we redefine $\phi$ as a module of
$\phi_{m^{\prime}}J_{m^{\prime}}(\xi)e^{i(m\theta-{\bf k}\cdot{\bf z})}$,
where $m^{\prime}=\pm m$, and $\phi_{m^{\prime}}$ is constant fixed later.
For $m^{\prime}=\pm 1/2$, note that $J_{-1/2}\gg J_{1/2}$
for $\mu^{\ast}\to 0$.
Wherein the Coulomb potential energy $V(R)$ can be obtained by applying
$\xi\coloneqq\mu^{\ast}r\bigl(=0^{(\prime)}\cdot\infty\bigr)\to
g_{0}^{-1}\bar{\mu}^{\ast}R$ to $\|\phi(m^{\prime}=-1/2)\|^{g_{0}}$,
that is, $\sim\phi_{-1/2}^{2}/2\xi\to\alpha/\bar{\mu}^{\ast}R
\eqqcolon V(R)/m_{e}c^{2}$, with $\phi_{-1/2}^{2}\to\alpha$ and
$\bar{\mu}^{\ast}=m_{e}c/\hbar$ (cf. Sect.~\ref{sec:3.2}).
Therefore, $\phi(m^{\prime}=-1/2)$ can be associated with internal
scalar field of the virtual photon concomitant with charged leptons;
the $\mu^{\ast}\to 0$ leading the indetermination,
with the massless state.

While the process (A$^{\prime}$) is requisite for the (B$^{\prime}$),
the latter is to concur with the former.
The problem herein is that the (B$^{\prime}$) should complete
the excitation of electron involving the connections among
its spatial motion, spin, and photon.
For this, there are well-established answers in the form based on
the Dirac and Maxwell/Proca wave equations~\cite{dirac1928}.
That is, these equations capture the connections in a way that is
impossible using scalar fields, leading to experimental predictions
that have precisely been verified.
This circumstance stimulates us to geometrically extend,
via the (B$^{\prime}$), the foregoing scalar form, reducing the
second order of derivative to the first one~\cite{waerden1974}.
The extension should be accomplished in such a way that the geometric
framework self-contains the scalar form reflected in the
lowest energy excitation of the elementary particles.
In the scalar one, $\phi$ satisfies
\begin{subequations}
\label{eq:4.3}
\begin{gather}
\left(\Delta+\kappa^{2}\right)\phi=0,
\label{eq:4.3a}
\\
\kappa^{2}={\bf k}^{2}+\mu^{\ast 2},
\label{eq:4.3b}
\end{gather}
\end{subequations}
\noindent
and \eqref{eq:4.2} with $\xi$ replacing $\mu r$ likewise.
Regarding the onset of $\hat{\bf z}$, here we introduce the
vector field ${\bf\Phi}$ transformed from $\phi$ as follows:
\begin{equation}
{\bf\Phi}({\bf r})=(\mu^{\ast})^{-2}
\left[\nabla\times\nabla\times\left(\phi\hat{\bf z}\right)
+\kappa\nabla\times\left(\phi\hat{\bf z}\right)\right].
\label{eq:4.4}
\end{equation}
\noindent
This is the solution of
\begin{equation}
\nabla\times\bf{\Phi}=\kappa\bf{\Phi},
\label{eq:4.5}
\end{equation}
\noindent
as far as $\phi$ satisfies \eqref{eq:4.3}~\cite{chandrasekhar1957}.
The field $\bf{\Phi}$ is identical with the Gromeka-Beltrami flow,
which is known as representation of force-free magnetic field
(proportional to current) of a plasma in ${\bf R}^{3}$ space.
It is noted that in the ideal magnetohydrodynamics in
which electric conductivity is assumed to be infinite, the form
as with \eqref{eq:4.5}, $\nabla\times\bf{B}=\lambda\bf{B}$ with
$\lambda$ being constant, has been derived from variation of the
magnetic energy $\mathscr{H}=\tfrac{1}{2}\int\|{\bf B}\|^{2}dV$
under the constraint that the magnetic helicity
$\mathscr{K}=\tfrac{1}{2}\int{\bf B}\cdot[(\nabla\times)^{-1}{\bf B}]dV$
is invariant, that is, $\delta\mathscr{H}=0=\delta\mathscr{H}
-\lambda\delta\mathscr{K}$~\cite{woltjer1958,taylor1986}.

Equation~\eqref{eq:4.5} can be recognized as the rotational eigenvalue
equation~\cite{moses1971}, so that $\bf{\Phi}$ is referred to as
rotational field.
In the case for which $m$ is arbitrary, the function of $\bf{\Phi}$
is written out below; this exemplifies generalization of the
special case $m=m^{\prime}=0$ or $\pm 1$ that the
classical field theory often refers to.
Substituting $\phi=\phi_{m^{\prime}}J_{m^{\prime}}(\xi)e^{i(m\theta-kz)}$
into \eqref{eq:4.4}, we obtain the expression of
${\bf\Phi}=(\Phi_{r},\Phi_{\theta},\Phi_{z})$, where
\begin{subequations}
\label{eq:4.6}
\begin{align}
\Phi_{r}&=i\frac{\phi_{m^{\prime}}}{\mu^{\ast}}\left[
\frac{m\kappa}{\xi}J_{m^{\prime}}(\xi)-k\frac{dJ_{m^{\prime}}(\xi)}{d\xi}
\right]e^{i\left(m\theta -kz\right)},
\label{eq:4.6a}
\\
\Phi_{\theta}&=\frac{\phi_{m^{\prime}}}{\mu^{\ast}}\left[
\frac{mk}{\xi}J_{m^{\prime}}(\xi)-\kappa\frac{dJ_{m^{\prime}}(\xi)}{d\xi}
\right]e^{i\left(m\theta -kz\right)},
\label{eq:4.6b}
\\
\Phi_{z}&=\phi_{m^{\prime}}J_{m^{\prime}}(\xi)e^{i\left(m\theta -kz\right)},
\label{eq:4.6c}
\end{align}
\end{subequations}
\noindent
and $\xi\notin\{0,\infty\}$.
The helical pattern of $\bf{\Phi}$ having rotational and translational
symmetry is characterized by $m$ as modal number and $k$
(e.g., see~\cite{sato1996} and the references, for the vivid
analogy of the classical field).
In particular, ${\bf\Phi}(m=\pm 1/2,m^{\prime}=\pm 1/2,k)$ can be
involved in \eqref{eq:4.2} via $\Phi_{z}=\phi$ with $\mu r$
replacing $\xi$ so that the ${\bf\Phi}$ with $\mu^{\ast}=\mu$
could serve as building blocks (internal structure, in part)
of spin $1/2$ point particles.
For an eigenflow of ${\bf\Phi}$, the corresponding eigenvalue of
$\xi$ is expected to indicate the mass as the Casimir invariant
of the Poincar{\'e} group.
As the observed mass results from the renormalization,
we have a reasoning that the geometric extension could
capture the interaction with covariant photons.

\subsection{Eigenflows of the Rotational Field}\label{sec:4.2}

Let us suppose interaction of the {\em single} particle with
gauge fields to be represented as the process in which the
internal structure of the particle refers to the eigenflows
of ${\bf\Phi}$, to be regulated.
In general, it is known that eigenvalues of the flows are not discrete,
spanning total complex plane, when the harmonic field is taken into
account~\cite{yoshida1990}.
The situation considered here falls under the case in which
$\kappa$ is real, provided $\xi$ and $k$ are real.
For seeking the specific solution, from the correspondence of
${\bf\Phi}$ to ${\bf B}$ follows: the eigenvalue problem of \eqref{eq:4.5}
could be formally compared to the aforementioned problem to
find the stationary point, at which ${\bf B}$-field configuration
minimizing $\mathscr{H}$ is characterized by the minimum
eigenvalue of $\lambda\,(=\!\!\mathscr{H}\!/\!\mathscr{K})$.
The nontrivial real-value of $\lambda$ is determined by a condition for
magnetic flux; the perfectly conducting wall confining the flux makes
the $\lambda$-value discrete for the cylindrical mode of $m=\pm 1$.
With these in mind, we suppose the internal Hamiltonian with
$\|{\bf\Phi}(m,m^{\prime})\|^{g_{0}}$ being the integrand, and that
the lowest energy excitation of the spinning particles refers to,
for the common $m$ and $m^{\prime}(=\pm m)$, the eigenvalue of
$\kappa$ whose magnitude is minimum (cf. the pion excitation
as the Nambu-Goldstone mode).
In determining the value, we impose on ${\bf\Phi}$ the boundary
condition of $\Phi_{r}(r=a)=0$, which can be expressed as
\begin{equation}
\frac{m\kappa}{\tilde{\xi}}J_{m^{\prime}}(\tilde{\xi})
-k\left[\frac{dJ_{m^{\prime}}(\xi)}{d\xi}\right]_{\xi=\tilde{\xi}}=0,
\label{eq:4.7}
\end{equation}
\noindent
where $\tilde{\xi}=\mu^{\ast}a$, and physical meaning of $a$ is
clarified later.
The root of equation~\eqref{eq:4.7} can provide an upper limit of $\xi$,
as the $\tilde{k}$ in Sect.~\ref{sec:3.1} does likewise to internally
regulate the orbital motion (confinement) of $q\bar{q}$
in the closed ${\bf R}^{3}$ space.
A well-defined domain can then be set in $\xi\gneq 0$, excluding
the infinity, as compatible with asymptotic property of the
harmonic field in the {\em internal} space.
The infinity is called for a specific purpose (cf. Sect.~\ref{sec:5.2.1}).

In the current context, it is natural to consider that the eigenflows of
${\bf\Phi}(m=\pm 1/2,m^{\prime})$ and ${\bf\Phi}(m=\pm 1,m^{\prime})$
involving discrete eigenvalues derived from \eqref{eq:4.7} determine
electromagnetic and weak susceptibility to the spin $1/2$ point particles
that are themselves with the non-discrete
${\bf\Phi}(m=\pm 1/2,m^{\prime})$ with $\mu^{\ast}=\mu$.
For example, the eigenflow of ${\bf\Phi}(m=m^{\prime}=-1/2)$ is
considered to internally regulate the orbital motion of a
charged $\ell_{\mathrm L}$ interacting with electromagnetic field,
in parallel to the foregoing quark picture.
In the following, we examine the eigenfunctions of
${\bf\Phi}(m=\pm 1/2,m^{\prime})$ for $m^{\prime}=\pm 1/2$ each.

\subsubsection{The Case of $m^{\prime}=1/2$}\label{sec:4.2.1}

Equation~\eqref{eq:4.7} for $(m,m^{\prime})=(1/2,1/2)$ can be expressed as
\begin{equation}
\tilde{\kappa}+\tilde{k}\left(1-2\tilde{\xi}\cot\tilde{\xi}\right)=0.
\label{eq:4.8}
\end{equation}
\noindent
Here, $\tilde{\kappa}=\kappa a$ and $\tilde{k}=ka$ are real, and we have
the relation of $\tilde{\kappa}=\pm\sqrt{\tilde{k}^{2}+\tilde{\xi}^{2}}$
from \eqref{eq:4.3b}.
Although the root property of \eqref{eq:4.7} depends on the sign of
$m\kappa$, rather than $\kappa$, the case analysis for the sign of
$\kappa$ each is carried out for given $m$, for an explanation.

\begin{figure}
\centerline{\includegraphics[width=10cm]{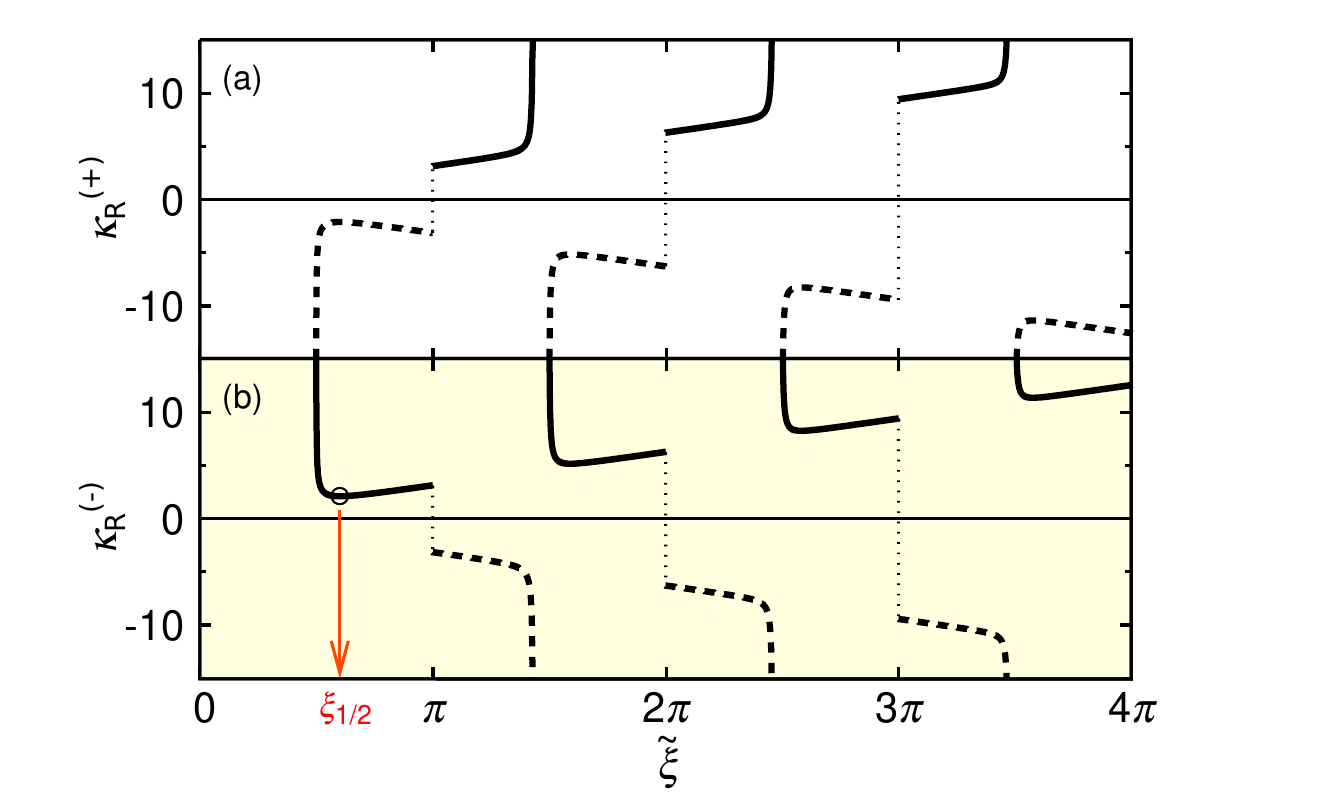}}%
\vspace{.1cm}
\caption{\label{fig:2}Plots of $\kappa_{\mathrm R}^{(+)}(\tilde{\xi})$
[equation~\eqref{eq:4.9}] (a) and $\kappa_{\mathrm R}^{(-)}(\tilde{\xi})$
[equation~\eqref{eq:4.10}] (b).
The functions positive in the discrete domains are shown by the
solid curves.
Both the figures have a common horizontal axis.
In~(b), the point at which $\kappa_{\mathrm R}^{(-)}$ takes the
minimum is indicated by the open circle.}
\end{figure}

\paragraph{For $\tilde{\kappa}>0$}
We express $\tilde{\kappa}$ as a function of $\tilde{\xi}$, eliminating
$\tilde{k}$, and write the positive $\tilde{\kappa}(\tilde{\xi})$ as
$\kappa_{\mathrm R}^{(+)}$, where the subscript $\mathrm R$
stands for right-handedness of the helix ($m>0$).
That is, we have
\begin{equation}
\kappa_{\mathrm R}^{(+)}
=\frac{\tilde{\xi}\left(2\tilde{\xi}\cot\tilde{\xi}-1\right)}
{\sqrt{\left(2\tilde{\xi}\cot\tilde{\xi}-1\right)^{2}-1}},
\label{eq:4.9}
\end{equation}
\noindent
valid for $\tilde{\xi}\cot\tilde{\xi}>1$.
In the discrete domains of definition, we investigate existence of
the minimum of $\kappa_{\mathrm R}^{(+)}$.
If there exists, we evaluate
$\kappa_{\mathrm R}^{(+)}=\tilde{\kappa}$, $\tilde\xi$, and
$\tilde{k}=\sqrt{\tilde{\kappa}^{2}-\tilde{\xi}^{2}}$
at the stable point: these constitute a set of the eigenvalues
for the eigenfunction of $\bf{\Phi}$.
Along this guideline, the function of
$\kappa_{\mathrm R}^{(+)}(\tilde{\xi})$ is plotted in Fig.~\ref{fig:2}(a).
It is noted that the root space should be restricted to a finite range,
as consistent with the finiteness of $\xi$.
Anyhow, we see that there is no local minimum in the domains.
It is thus concluded that for $(m,m^{\prime},\kappa)=(1/2,1/2,+)$
exists no eigenstate.

\paragraph{For $\tilde{\kappa}<0$\,}
For \eqref{eq:4.8}, $-\tilde{\kappa}$ is expressed as a function of
$\tilde{\xi}$, and the positive $-\tilde{\kappa}(\tilde{\xi})$ is
denoted as $\kappa_{\mathrm R}^{(-)}$; that is,
\begin{equation}
\kappa_{\mathrm R}^{(-)}=-\kappa_{\mathrm R}^{(+)},
\label{eq:4.10}
\end{equation}
\noindent
for $\cot\tilde{\xi}<0$.
In Fig.~\ref{fig:2}(b), the function of $\kappa_{\mathrm R}^{(-)}$
is plotted; now we find the local minima in the discrete domains.
The equality $d\kappa_{\mathrm R}^{(-)}(\tilde\xi)/d\tilde\xi =0$
provides the eigenvalue equation for $\tilde\xi$,
which can be expressed as
\begin{equation}
\label{eq:4.11}
\left(\tau -4\upsilon^{-1}\right)\left(\tau -\upsilon^{-1}\right)
\upsilon^{-1}+\tau=0,
\end{equation}
\noindent
for $\upsilon^{-1}<0$, where
$\tau={\tilde\xi}^{-1}$ and $\upsilon=\tan{\tilde\xi}$.
The minimum root of \eqref{eq:4.11} is denoted as $\xi_{1/2}$,
where the subscript indicates the $m^{\prime}$-value.
Choosing the minimum value among the local minimum values
guarantees the foregoing restriction of root space.
It follows that the well-defined
$\kappa_{1/2}=\kappa_{\mathrm R}^{(-)}(\tilde\xi =\xi_{1/2})$ and
$k_{1/2}=\sqrt{\kappa_{1/2}^{2}-\xi_{1/2}^{2}}$ are evaluated, to give
the rotational eigenvalue of $-\kappa_{1/2}(\xi_{1/2},k_{1/2})$.
This means that, even for the previous case in which $\kappa>0$,
we could have the eigenvalue of $\kappa_{1/2}(\xi_{1/2},k_{1/2})$
by the change of $m=1/2\to -1/2$, i.e.,
$(m,m^{\prime},\kappa)=(-1/2,1/2,+)$.
The values of $\xi_{1/2}$, $\kappa_{1/2}$, and $k_{1/2}$ are listed
in the upper row of Table~\ref{tb:1}.

\begin{table}[b]
\begin{center}
\begin{minipage}{160pt}
\caption{A set of the discrete eigenvalues
$\xi_{m^{\prime}}$, $\kappa_{m^{\prime}}$, and $k_{m^{\prime}}$
for $m^{\prime}=\pm 1/2$.}\label{tb:1}%
\begin{tabular}{@{}cccc@{}}
\toprule
$m^{\prime}$ & $\xi_{m^{\prime}}$  & $\kappa_{m^{\prime}}$ &
$k_{m^{\prime}}$\\
\midrule
$\hphantom{-}1/2$ & $1.891$ & $2.110$ & $0.9363$ \\
$-1/2$ & $3.445$ & $3.632$ & $1.152\hphantom{0}$ \\
\botrule
\end{tabular}
\end{minipage}
\end{center}
\end{table}

\subsubsection{The Case of $m^{\prime}=-1/2$}\label{sec:4.2.2}

Equation~\eqref{eq:4.7} for $(m,m^{\prime})=(-1/2,-1/2)$ can be
expressed as
\begin{equation}
\tilde{\kappa}-\tilde{k}\left(1+2\tilde{\xi}\tan\tilde{\xi}\right)=0.
\label{eq:4.12}
\end{equation}
\noindent
By the method explained above, we look for the eigenvalues
the \eqref{eq:4.12} contains.

\begin{figure}
\centerline{\includegraphics[width=10cm]{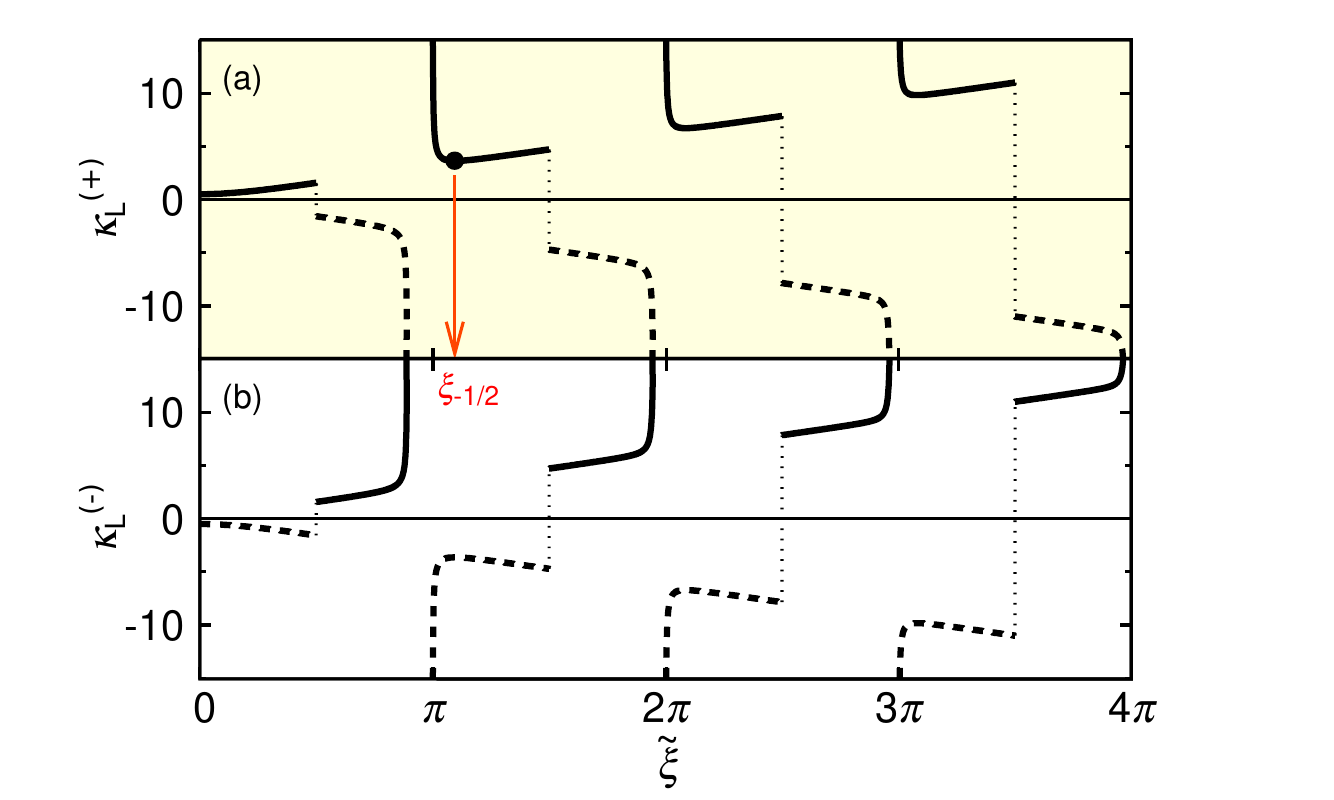}}%
\vspace{.1cm}
\caption{\label{fig:3}Plots of $\kappa_{\mathrm L}^{(+)}(\tilde{\xi})$
[equation~\eqref{eq:4.13}] (a) and $\kappa_{\mathrm L}^{(-)}(\tilde{\xi})$ (b).
The functions positive in the discrete domains are shown by
the solid curves.
Both the figures have a common horizontal axis.
In~(a), the point at which $\kappa_{\mathrm L}^{(+)}$ takes the minimum
is indicated by the filled circle.}
\end{figure}

\paragraph{For $\tilde{\kappa}>0$\,}
For \eqref{eq:4.12}, we express $\tilde{\kappa}$ as a function of
$\tilde{\xi}$, eliminating $\tilde{k}$, and write the positive
$\tilde{\kappa}(\tilde{\xi})$ as $\kappa_{\mathrm L}^{(+)}$, where the
subscript $\mathrm L$ stands for left-handedness of the helix ($m<0$).
That is, we have
\begin{equation}
\kappa_{\mathrm L}^{(+)}
=\frac{\tilde{\xi}\left(2\tilde{\xi}\tan\tilde{\xi}+1\right)}
{\sqrt{\left(2\tilde{\xi}\tan\tilde{\xi}+1\right)^{2}-1}},
\label{eq:4.13}
\end{equation}
\noindent
valid for $\tan\tilde{\xi}>0$.
If there exists the minimum of $\kappa_{\mathrm L}^{(+)}$
in the domain of definition,
$\kappa_{\mathrm L}^{(+)}=\tilde\kappa$, $\tilde\xi$, and $\tilde k$
are evaluated at the stable point.
In Fig.~\ref{fig:3}(a), the function of $\kappa_{\mathrm L}^{(+)}$ is plotted,
to show the profile similar to that displayed in Fig.~\ref{fig:2}(b),
and existence of the local minima in the discrete domains.
The equality $d\kappa_{\mathrm L}^{(+)}(\tilde\xi)/d\tilde\xi =0$ provides
\begin{equation}
\left(\tau +4\upsilon\right)\left(\tau +\upsilon\right)\upsilon-\tau=0,
\label{eq:4.14}
\end{equation}
for $\upsilon >0$.
The minimum root of \eqref{eq:4.14} is denoted as
$\tilde\xi =\xi_{-1/2}$, which determines the values of
$\kappa_{-1/2}=\kappa_{\mathrm L}^{(+)}(\tilde\xi =\xi_{-1/2})$
and $k_{-1/2}=\sqrt{\kappa_{-1/2}^{2}-\xi_{-1/2}^{2}}$,
to give $\kappa_{-1/2}(\xi_{-1/2},k_{-1/2})$.
These values are listed in the bottom row of Table~\ref{tb:1}.

\paragraph{For $\tilde{\kappa}<0$\,}
For \eqref{eq:4.12}, $-\tilde{\kappa}$ is expressed as a
function of $\tilde{\xi}$, and the positive $-\tilde{\kappa}(\tilde{\xi})$ is
denoted as $\kappa_{\mathrm L}^{(-)}$.
The function of $\kappa_{\mathrm L}^{(-)}=-\kappa_{\mathrm L}^{(+)}$,
valid for $\tilde\xi\tan\tilde\xi<-1$, is plotted in Fig.~\ref{fig:3}(b).
As expected, there is no local minimum in the domains:
no eigenstate in this case.
And yet, by the change of $m=-1/2\to 1/2$, i.e.,
$(m,m^{\prime},\kappa)=(1/2,-1/2,-)$, we could have
$-\kappa_{-1/2}(\xi_{-1/2},k_{-1/2})$.

\subsection{Chiral Asymmetry of the Helical Eigenflows and
a Representation of Beta-Decay Products}\label{sec:4.3}

In light of the allowed combinations of $(m,m^{\prime},\kappa)=(\pm,\pm,\pm)$,
the rotational eigenvalues for $m=\pm 1/2$ are summarized in Table~\ref{tb:2}.
It is proclaimed that the eigenstates do exist for $m\kappa <0$,
whereas do not for $m\kappa >0$.
For the former, $\Phi_{z}$ of the eigenflows for
$(m,m^{\prime},\kappa)=(-,-,+)$, $(+,+,-)$, $(+,-,-)$, and $(-,+,+)$
responds to the term of $J_{-1/2}e^{-i\theta/2}$ in \eqref{eq:4.2a},
$iJ_{1/2}e^{i\theta/2}$ \eqref{eq:4.2b},
$J_{-1/2}e^{i\theta/2}$ \eqref{eq:4.2c},
and $-iJ_{1/2}e^{-i\theta/2}$ \eqref{eq:4.2d}, respectively,
with $\xi=\mu r$ on the common cylindrical base.
For the latter, $\Phi_{z}$ of the non-eigenflows for
$(m,m^{\prime},\kappa)=(-,-,-)$, $(+,+,+)$, $(+,-,+)$, and $(-,+,-)$
responds to $J_{-1/2}e^{-i\theta/2}$ in \eqref{eq:4.2d},
$-iJ_{1/2}e^{i\theta/2}$ \eqref{eq:4.2c},
$J_{-1/2}e^{i\theta/2}$ \eqref{eq:4.2b},
and $iJ_{1/2}e^{-i\theta/2}$ \eqref{eq:4.2a}, respectively.
Lepton doublet transition in the $\beta^{-}$-decay:
$(\nu_{e})_{\mathrm L}\to e^{-}_{\mathrm L}$ can be compared to
transition of $\kappa_{1/2}\to\kappa_{-1/2}$ for $(m,\kappa)=(-,+)$, while
$(\bar{\nu}_{e})_{\mathrm R}\to e^{+}_{\mathrm R}$ in the $\beta^{+}$-decay,
to $-\kappa_{1/2}\to-\kappa_{-1/2}$ for $(m,\kappa)=(+,-)$.
The $CP$ transformation can be seen as the flip of
$(m,\kappa)=(-,+)\rightleftarrows(+,-)$ for $m^{\prime}$ unchanged.

\begin{table}[t]
\begin{center}
\begin{minipage}{190pt}
\caption{Summary of the rotational eigenvalues.}\label{tb:2}%
\begin{tabular}{@{}ccccc@{}}
\toprule
$m$ & $m^{\prime}$ & $\kappa\,(>0)$ & $\kappa\,(<0)$ &
reference\footnotemark[3]\\
\midrule
$m^{\prime}\,({\mathrm{L}})$ & $-1/2$ & $\kappa_{-1/2}$ & --- &
\eqref{eq:4.2a}\\
$m^{\prime}\,({\mathrm{R}})$ & $+1/2$ & --- & $-\kappa_{1/2}$ &
\eqref{eq:4.2b}\\
$-m^{\prime}\,({\mathrm{R}})$ & $-1/2$ & --- & $-\kappa_{-1/2}$
& \eqref{eq:4.2c}\\
$-m^{\prime}\,({\mathrm{L}})$ & $+1/2$ & $\kappa_{1/2}$ & --- &
\eqref{eq:4.2d}\\
\botrule
\end{tabular}
\footnotetext[3]{Equation numbers for the leptonic states that
refer to the eigenvalues.}
\end{minipage}
\end{center}
\end{table}

We now see that, for the reaction~\eqref{eq:3.1}, $\ell_{\mathrm L}$ and
$(\bar{\ell})_{\mathrm R}$ must internally refer to the eigenflows of
${\bf\Phi}(m=-1/2,m^{\prime}=\mp 1/2,\tilde\kappa=\kappa_{\mp 1/2})$
and ${\bf\Phi}(1/2,\pm 1/2,-\kappa_{\pm 1/2})$, respectively;
in particular, charged $\ell_{\mathrm L}$ and neutral
$(\bar{\ell})_{\mathrm R}$, to ${\bf\Phi}(m=m^{\prime}=-1/2,\kappa_{-1/2})$
and ${\bf\Phi}(m=m^{\prime}=1/2,-\kappa_{1/2})$, respectively.
Noteworthy is the relation of $\kappa_{-1/2}\neq\kappa_{1/2}$,
and chiral asymmetry between L- and R-helix of the eigenflows.
This asymmetry is amenable to left-handed selectivity of the charged
current weak interaction.
According to these ingredients, I put forth the module representation of
\begin{subequations}
\label{eq:4.15}
\begin{align}
&{\bf\Phi}\left(-\tfrac{1}{2},-\tfrac{1}{2},+\right)
+i{\bf\Phi}\left(-\tfrac{1}{2},\tfrac{1}{2},-\right)
\eqqcolon{\bf\Psi}_{-},
\label{eq:4.15a}\\
&{\bf\Phi}\left(\tfrac{1}{2},-\tfrac{1}{2},+\right)
+i{\bf\Phi}\left(\tfrac{1}{2},\tfrac{1}{2},-\right)
\eqqcolon{\bf\Psi}_{+},
\label{eq:4.15b}
\end{align}
\end{subequations}
\noindent
for respectively the charged $\ell_{\mathrm L}$ and neutral
$(\bar{\ell})_{\mathrm R}$ created in the~\eqref{eq:3.1}.
Here, the second and first term in \eqref{eq:4.15a} and \eqref{eq:4.15b},
respectively, stand for the orthogonal counterpart having the inverted
$m^{\prime}$ and $\kappa$, with respect to
${\bf\Phi}(m,m^{\prime},\kappa)$, so that $m\kappa>0$.
The expression of ${\bf\Psi}_{\mp}$, i.e., the superposition of the
non-discrete flows with $\mu^{\ast}=\mu$ can suitably configure
\eqref{eq:4.2a} and \eqref{eq:4.2b}, respectively.
Provided the indication of the lepton mass of
$\xi_{\mp 1/2}\to m_{\ell,\bar{\ell}}$ (cf. Sect.~\ref{sec:6.2}),
${\bf\Psi}_{\mp}(\mu,k,\|m\|)$ can be compared to the irreducible
representation of the Poincar{\'e} group~\cite{wigner1939}: 
the free-particle states $\lvert\left. m_{\ell,\bar{\ell}},p,s\right>$,
where $p=\|{\bf P}\|$, and $s=1/2$ denotes the spin quantum
number in ${\bf S}^{2}=s(s+1)\hbar^{2}$.

\subsection{Remarks on the Integer Modes}\label{sec:4.4}

For $m=\pm 1$, we outline the results, in that the eigenflows are
related to gauge bosons with the spin $\|m\|$.
One can calculate the discrete eigenvalues basically along the
procedure described above.
However, we have only to examine the case of $m=m^{\prime}$,
as $J_{m^{\prime}}$ and $J_{-m^{\prime}}$ with $m^{\prime}$
integer are linearly dependent.

In general, on pattern of existence of the eigenstates,
it is found that $\kappa_{\mathrm R}^{(-)}(=-\tilde\kappa >0)$
and $\kappa_{\mathrm L}^{(+)}(=\tilde\kappa >0)$ with
respectively $(m,\kappa)=(+,-)$ and $(-,+)$, i.e.,
$m\kappa <0$, have local minima in their domains.
As is, a triad of the well-defined eigenvalues
$\xi_{m}$, $\kappa_{m}$, and $k_{m}$ can be obtained.
For the mode in which $m$ is an even number, we have
$\xi_{m}=\xi_{-m}$, $\kappa_{m}=\kappa_{-m}$, and $k_{m}=k_{-m}$,
whereas the odd mode provides $\xi_{\|m\|}<\xi_{-\|m\|}$,
$\kappa_{\|m\|}<\kappa_{-\|m\|}$, and $k_{\|m\|}<k_{-\|m\|}$
(like for $m=\pm 1/2$), breaking chiral symmetry.
This modal parity stems from the relation of $J_{-m}=(-1)^{m}J_{m}$.
In the special case of $m=0$, spectrum of $\tilde{\kappa}$ is continuous.

\begin{figure}
\centerline{\includegraphics[width=10cm]{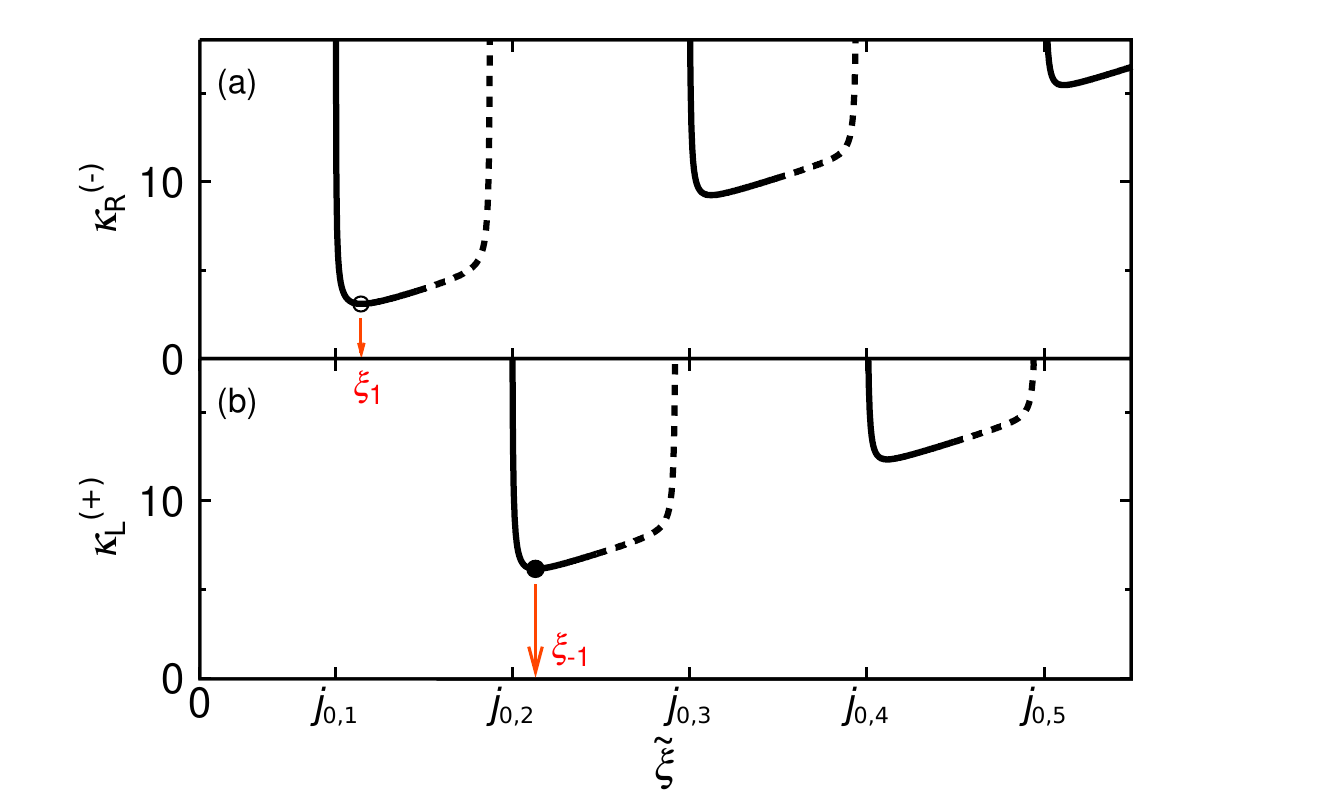}}%
\vspace{.1cm}
\caption{\label{fig:4}Plots of $\kappa_{\mathrm R}^{(-)}(\tilde{\xi})$
[equation~\eqref{eq:4.16}] (a) and
$\kappa_{\mathrm L}^{(+)}(\tilde{\xi})
=-\kappa_{\mathrm R}^{(-)}(\tilde{\xi})$ (b).
These functions having, respectively, the domain of
$j_{0,n}<\tilde\xi <j_{1,n}$ for odd and even number of
$n\,(=1,2,3,\dots)$ are shown by the solid curves.
Both the figures have a common horizontal axis, on which positions
of $j_{0,n}$ are indicated, in place of scale of the linear axis.
The points at which $\kappa_{\mathrm R}^{(-)}$ and
$\kappa_{\mathrm L}^{(+)}$ take the minima in their domains are
indicated by the open and the filled circle, respectively.
The broken curves in~(a,b) correspond to
$\kappa_{\mathrm L}^{(-)}$ and $\kappa_{\mathrm R}^{(+)}$ for
$j_{1,n}<\tilde\xi <j_{2,n}$ with $n$ odd and even, respectively (see text).}
\end{figure}

For $m=\pm 1$ of particular interest, the function of
$\kappa_{\mathrm R}^{(-)}(\tilde\xi)$ is expressed as
\begin{equation}
\kappa_{\mathrm R}^{(-)}=
-\frac{{\tilde\xi}\left[{\tilde\xi}J_{0}({\tilde\xi})
-J_{1}({\tilde\xi})\right]}
{\sqrt{{\tilde\xi}J_{0}({\tilde\xi})\left[{\tilde\xi}J_{0}({\tilde\xi})
-2J_{1}({\tilde\xi})\right]}},
\label{eq:4.16}
\end{equation}
\noindent
which is related to $\kappa_{\mathrm L}^{(+)}(\tilde\xi)$ as
$-\kappa_{\mathrm R}^{(-)}=\kappa_{\mathrm L}^{(+)}$.
Here, the domains of $\kappa_{\mathrm R}^{(-)}$ and
$\kappa_{\mathrm L}^{(+)}$ positive definite are given by
$j_{0,n}<\tilde\xi <j_{1,n}$ with $n$ being odd and even number,
respectively, where $j_{m,n}$ denotes the $n$-th zero of
$J_{m}(\tilde\xi)$.
The $\kappa_{\mathrm R}^{(-)}$ and $\kappa_{\mathrm L}^{(+)}$ are plotted
in Fig.~\ref{fig:4}(a,b), respectively.
We find the local minima in the domains, and that
$\kappa_{\mathrm L}^{(-)}$ and $\kappa_{\mathrm R}^{(+)}$
with $m\kappa>0$ have no such points in their domains given by
$j_{1,n}<\tilde\xi <j_{2,n}$ with $n$ odd and even, respectively.
The eigenvalue equation for $\tilde\xi$ can be written as
\begin{equation}
J_{0}^{2}({\tilde\xi})\left[{\tilde\xi}J_{0}({\tilde\xi})
-2J_{1}({\tilde\xi})\right] +J_{1}^{3}({\tilde\xi})=0.
\label{eq:4.17}
\end{equation}
\noindent
The roots in $j_{0,1}<\tilde\xi <j_{1,1}$ and $j_{0,2}<\tilde\xi <j_{1,2}$
are denoted as $\xi_{1}$ and $\xi_{-1}$, respectively; they provide
$\kappa_{1}=\kappa_{\mathrm R}^{(-)}({\tilde\xi}=\xi_{1})$,
$\kappa_{-1}=\kappa_{\mathrm L}^{(+)}({\tilde\xi}=\xi_{-1})$,
and $k_{\pm 1}=\sqrt{\kappa_{\pm 1}^{2}-\xi_{\pm 1}^{2}}$.
For convenience, the values are listed in Table~\ref{tb:3}.\\

\begin{table}[!h]
\begin{center}
\begin{minipage}{160pt}
\caption{A set of the discrete eigenvalues $\xi_{m}$, $\kappa_{m}$, and
$k_{m}$ for $m=\pm 1$.}\label{tb:3}%
\begin{tabular}{@{}cccc@{}}
\toprule
$m\,(=m^{\prime})$ & $\xi_{m}$ & $\kappa_{m}$ & $k_{m}$ \\
\midrule
${\hphantom{-}}1$ & $2.857$ & $3.112$ & $1.234$ \\
$-1$ & $5.937$ & $6.162$ & $1.652$ \\
\botrule
\end{tabular}
\end{minipage}
\end{center}
\end{table}

\section{Observability of the Rotational Field}\label{sec:5}

\subsection{Relation of the Field with Nonrelativistic Kinematics of a
Fermionic Particle}\label{sec:5.1}

We consider application of the developed formalism to an observational
issue on ${\bf\Psi}_{-}$ including the interactive module
${\bf\Phi}(m=m^{\prime}=-1/2,\kappa>0)$.
Concerning the operation with $r\to\infty$, we do not have the
constraint on $k\to 0$ that has been required for the
$\|m\|=1/2$ system of internal pion field to expand
as compatible with the $\|m\|=1$ system.
This resembles the situation in which the transformation \eqref{eq:3.8}
is applicable to reproduction of the Cornell potential for
$q\bar{q}$ moving nonrelativistically.
It is, therefore, conceivable that kinematics derived from the
${\bf\Psi}_{-}$ for $k\to 0$ is of the charged $\ell$ moving
nonrelativistically (i.e., $\beta\to 0$ and $\bar{\mu}\neq 0$).
When $k\to 0$ is taken for $k\ne 0$ prerequisite to the reference to
an eigenflow, the operation can be translated as projection of the
helical field onto the cylindrical base.
For spatial transformation reflecting it, we invoke the extended form of
\begin{equation}
\kappa{\bf r}\,\longrightarrow\mathrm{sgn}(\kappa)\,g_{0}^{-1}\mu{\bm\rho},
\label{eq:5.1}
\end{equation}
\noindent
where $\mathrm{sgn}(\kappa\gtrless 0)=\pm 1$, ${\bm\rho}$ is radial
coordinate vector on a complex plane (with its origin $o$), and
$\mu\|{\bm\rho}\|=\mu r$.

For kinematic states of a single electron, we see the geometric
mechanics that covers $\mathrm{SU(2)}$, say, the Lie algebraic
representation of ${\bf S}=(\hbar/2)\bm{\sigma}$, where
$\bm{\sigma}$ is the vector of Pauli matrices:
$(\sigma_{1},\sigma_{2},\sigma_{3})$~\cite{pauli1927}.
In the special case of $k=0$ compared to the rest state, it is trivial
that sign of ${\bf\Psi}_{-}$ is inverted by the rotation of
$\theta=0\to 2\pi$, and recovered first by that of $4\pi$.
We render the rotation center $o$ identical with the origin $O$
at which the electron is at rest.
When making the rotation angle of either $-\theta$ or $\theta$
correspond to ${\bf R}^{2}$-rotation angle $\Theta$, therefore,
${\bf\Psi}_{-}$ captures a basic property of $\bf S$ that rotational
operator generates $\exp\left[i{\bm\Theta}\cdot{\bf S}/\hbar
\right]_{\|{\bm\Theta}\|=2\pi}=-I$, where $\|{\bm\Theta}\|=\Theta$,
and $I$ ($=\sigma_{1}^{2}=\sigma_{2}^{2}=\sigma_{3}^{2}$) is the
unit matrix.

\begin{figure}
\centerline{\includegraphics[trim=50 162 100 170,width=11cm,clip]{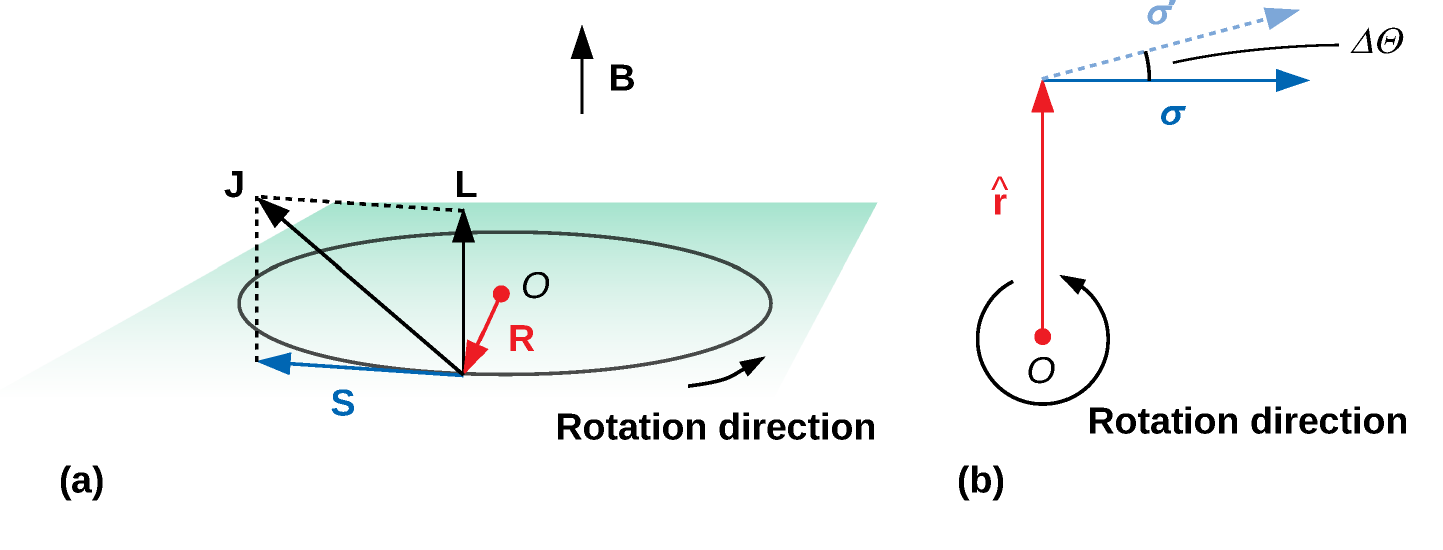}}%
\caption{\label{fig:5}Cyclotron motion of a single electron having
the spin ${\bf S}$ and its longitudinally polarized state (a),
and the corresponding rotation of unit radial vector
$\hat{\bf r}={\bf R}/\|{\bf R}\|$ and change of the polarization for
the one orbital turn: $\bm{\sigma}\to\bm{\sigma}^{\prime}$(b).
In~(a), the angular momentum coupling is depicted.}
\end{figure}

For the case of $k\neq 0$, we take account of motion of spin and
circular orbit of an electron affected by time-independent uniform
magnetic field ${\bf B}$, as shown in Fig.~\ref{fig:5}(a) (see figure~3-3
in \cite{sakurai1967}, and the relevant explanations therein).
The given is a simple situation in which the helicity is
conserved, according to the original Dirac's theory:
relativistic quantum mechanics.
Concerning access to ${\bf\Phi}$ of the nonrelativistic electron,
transformation of $kz\to g_{0}^{-1}(1-\delta_{g})\bar{\mu}\zeta$,
in conjunction with \eqref{eq:5.1}, is applied to the phase factor of
$\Phi_{z}$ raised to the power of $g_{0}$.
Here, $\mu$ has been replaced by $(1-\delta_{g})\bar{\mu}$,
with $\delta_{g}$ being $1-\sqrt{1-\delta}=\delta/2+\cdots$.
On the resulting $e^{-i[\theta+(1-\delta_{g})\bar{\mu}\zeta]}$, we impose
the lifting shift of $(\theta,\zeta)=(0,0)\to (-2\pi,2\pi\bar{\mu}^{-1})$,
which advances the phase.
In accordance with the L/R definition for helix, let negative sign
of rotational angles be clockwise.
Denoting by $\vD\theta$ the advanced phase, we get
$\delta_{g}=\vD\theta/2\pi$, which can be compared to
radiative correction term in the $g$-factor of
$2\left[1+\vD\Theta/(2\pi\Gamma)\right]$, where
$\Gamma=(1-\beta^{2})^{-1/2}$, and
$\vD\Theta=\Gamma\vD\Theta(\Gamma\to 1)$
is the observed advancing angle of spin precession, namely,
the Thomas precession.
For anti-clockwise, $\Theta(=-\theta)=2\pi$ rotation of unit coordinate
vector $\hat{\bf r}$ as shown in Fig.~\ref{fig:5}(b), we find the
correspondence between $\vD\Theta/\Gamma$ and $\vD\theta$:
\begin{equation}
\vD\theta\,\longleftrightarrow\,\vD\Theta(\beta\to 0).
\label{eq:5.2}
\end{equation}
\noindent
Our approach allows for the quantum correction of the pure Dirac particle
for which $\vD\Theta=0$.

\subsection{Internal Mechanism Advancing Spin Precession
Frequency}\label{sec:5.2}

\subsubsection{Coordinate Mechanics of the Charged Fermion}\label{sec:5.2.1}

The internal mechanics determining the value of $\vD\theta$ is to
respond to the full information of the observed kinematic state,
which is contained in the total angular momentum as the coupling of
the orbital angular momentum $\bf L$ and $\bf S$ (e.g., \cite{bohm1951}):
\begin{subequations}
\label{eq:5.3}
\begin{equation}
{\bf J}={\bf L}+{\bf S},
\label{eq:5.3a}
\end{equation}
as depicted in Fig.~\ref{fig:5}(a).
We pay attention to the formalism in which the \eqref{eq:5.3a}
divided by $\hbar$, for $\beta\to 0$, gives
\begin{gather}
{\bf J}/\hbar\,\longrightarrow\,{\bf S}/\hbar.
\label{eq:5.3b}
\end{gather}
\end{subequations}
\noindent
Reminding this, we find out internal coordinates reflected in the
rotational $\hat{\bf r}$ and $\bm{\sigma}$.
By combining the correspondence ${\bf\Phi}\leftrightarrow{\bf B}$
with \eqref{eq:2.3}, we prepare the function-space transformation of
${\bf\Phi}\to i{\tilde\lambda}{\bf r}$, where $\tilde\lambda$ is
dimensionless real-constant, for its application to equation
\eqref{eq:4.5} with $\mu$ replacing $\mu^{\ast}$.
Taking this into account, we introduce the covering vector
for ${\bf J}/(\hbar\surd{3})$, defined by
\begin{equation}
{\bf\Lambda}\coloneqq\left[\nabla\times{\bf\Phi}
\right]_{{\bf\Phi}\to i{\tilde\lambda}{\bf r}}=i{\tilde\lambda}\kappa{\bf r}.
\label{eq:5.4}
\end{equation}
\noindent
The terminology of the internal vector is from interpreting
the helical ${\bf\Phi}$ as a covering space.
Now, the transformation~\eqref{eq:5.1} cooperated with $k\to 0$
is applied to ${\bf\Lambda}$.
We then have the following form just in parallel with
\eqref{eq:5.3b}:
\begin{subequations}
\label{eq:5.5}
\begin{equation}
{\bf\Lambda}\,\longrightarrow\,{\bf\Lambda}_{0},
\label{eq:5.5a}
\end{equation}
where
\begin{gather}
{\bf\Lambda}_{0}=\mathrm{sgn}(\kappa)\,i{\tilde\lambda}g_{0}^{-1}\mu\bm{\rho}.
\label{eq:5.5b}
\end{gather}
\end{subequations}
\noindent
In the context, the quantity ${\bf\Lambda}_{0}$ with
$\mathrm{sgn}(\kappa)=+1$ must describe ${\bf S}/(\hbar\surd{3})$
of the electron moving nonrelativistically;
the condition is investigated below.
At the outset, we normalize \eqref{eq:5.5b}, to provide the following
symbolic master equation that spin-half particles are to refer to:
\begin{subequations}
\label{eq:5.6}
\begin{equation}
\hat{\sigma}=\mathrm{sgn}(\kappa)\,i\xi\hat{\rho},
\label{eq:5.6a}
\end{equation}
where
\begin{gather}
\hat{\sigma}={\bf\Lambda}_{0}/(\tilde{\lambda}/g_{0}),\mspace{50mu}
\hat{\rho}=\bm{\rho}/\|\bm{\rho}\|,
\label{eq:5.6b}
\end{gather}
\end{subequations}
\noindent
and $\xi=\mu\|\bm{\rho}\|$, which connotes coupling of
$\mu\|\bm{\rho}\|$ with $\mu^{\ast}r\eqqcolon\xi$
(Sect.~\ref{sec:4.1}).
Equation~\eqref{eq:5.6a} captures orthogonality of the base spaces
accommodating $(\bm{\rho},{\bf\Lambda}_{0})$,
to serve as the coordinate-rotor.
This is expected to rule infinitesimal rotations (in the sense of
standard analysis), as intuitively seen by the familiar form:
the isomorphism between $\mathrm{SU(2)}$ and $\mathrm{SO(3)}$
Lie algebra with the generators $S_{i}$ and $X_{i}$, respectively,
is sustained via $S_{i}=i\chi X_{i}$, where $\chi$ is constant
that depends on the way of defining the basis vectors.

\begin{figure}
\centerline{\includegraphics[trim=55 120 60 110,width=11cm,clip]{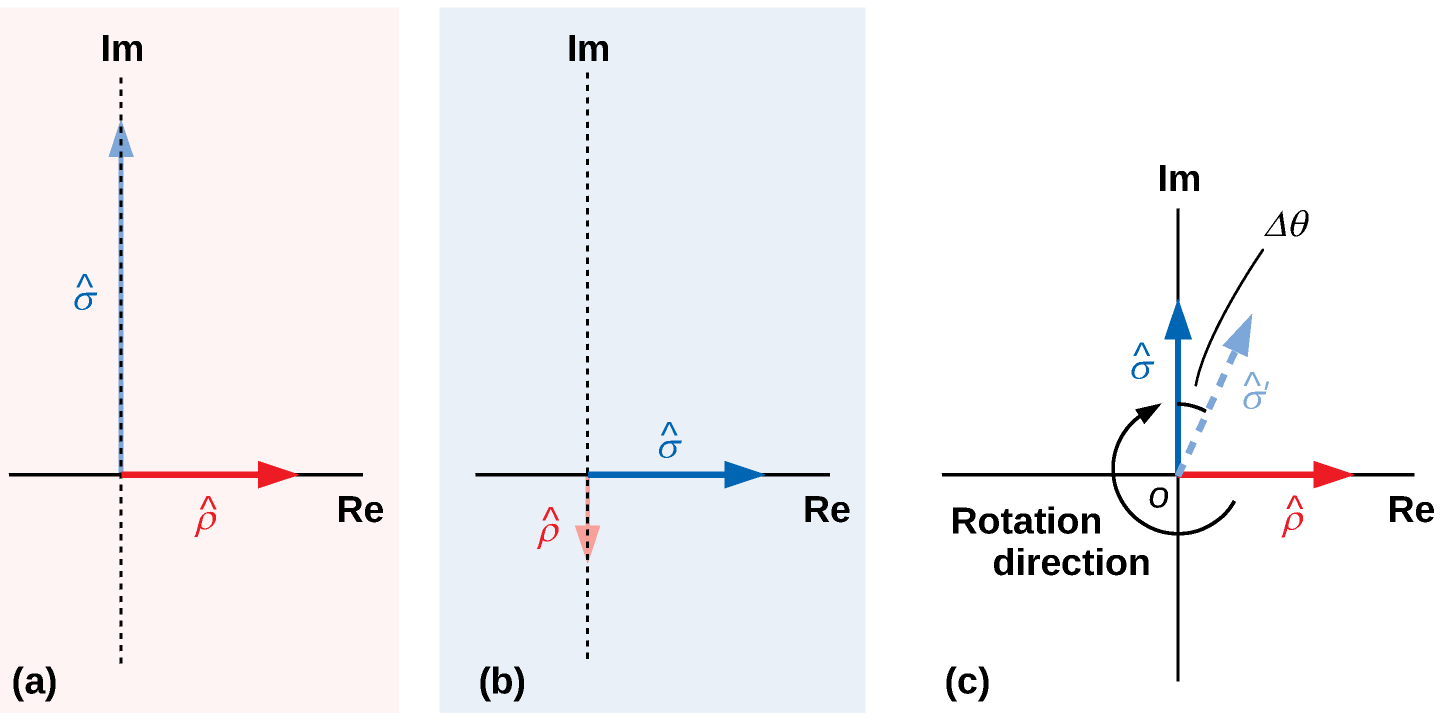}}%
\caption{\label{fig:6}Complex planes setting, on the real axes,
$\|\hat{\rho}\|=1$ (a) and $\|\hat{\sigma}\|=1$ (b) for a common
$\xi$-value $(>1)$, and a normalized complex plane (c) corresponding to
Fig.~\ref{fig:5}(b).}
\end{figure}

We delve into realistic rotation of the internal $(\bm{\rho},{\bf\Lambda}_{0})$
for $\mathrm{sgn}(\kappa)=+1$, in terms of
self-consistent procedure of the normalization.
First, suppose that $\hat\rho$ is a basis vector normalized when
$\bm{\rho}$ gets on real axis of the complex plane.
Then, trivially $\|\hat{\rho}\|=1$ is set on the real axis,
and the real basis vector is to come out in $\hat{\bf r}$.
The vector configuration of \eqref{eq:5.6a} on this plane is illustrated
in Fig.~\ref{fig:6}(a), providing a value of $\xi$ larger than unity.
Second, suppose that $\hat{\sigma}$ is another independent
basis vector normalized when $\bm{\rho}$ rotates by $-\pi/2$
to be purely imaginary and ${\bf\Lambda}_{0}$ gets on the real axis.
For ${\tilde\lambda}=1$, we can identify
$\|{\bf\Lambda}_{0}\|={\tilde\lambda}/g_{0}$ with
$\|{\bf S}\|/(\hbar\surd{3})=1/2$.
Then, $\|\hat{\sigma}\|=1$ is set on the real axis;
the vector configuration of \eqref{eq:5.6a} on the second plane is
shown in Fig.~\ref{fig:6}(b).
Besides, the vector configuration of $(\hat{\bf r},\bm{\sigma})$
[Fig.~\ref{fig:5}(b)] is recast to the one of $(\hat{\rho},\hat{\sigma})$
on a normalized complex plane, which is shown in Fig.~\ref{fig:6}(c).
This can be regarded as superposition of $\hat\rho$ on the
first plane (a) and $\hat\sigma$ put on the imaginary axis
by the rotation of $\pi/2$ on the second plane (b).
For $\Theta$ and $\theta$ as rotation angles of $\hat{\bf r}$ and
$\hat\rho$, respectively, we have $\Theta=-\theta$, and therefore,
$\hat\rho$ rotated by $-2\pi$ is to coincide with the initial one.
On the other hand, $\bm{\sigma}\to\bm{\sigma}^{\prime}$ incidental to
the one turn of $\hat{\bf r}$ should be described by
$\hat{\sigma}\to\hat{\sigma}^{\prime}$ with $\vD\theta$,
where $\|\hat{\sigma}\|=\|\hat{\sigma}^{\prime}\|=1$ is imposed.
Then, $\hat{\sigma}$ captures $\bm{\sigma}/\surd{3}$,
on account of $\hat{\sigma}^{2}\to\bm{\sigma}^{2}/3=I$.
That tips of both the $\hat{\rho}$ and $\hat{\sigma}$ sweep
unit circle on the plane (c) represents unitary evolution of
${\bf\Psi}_{-}$, and is necessary for the normalized
two-dimensional complex space to be the point:
$\hat{\rho}+i\hat{\sigma}=0$ (cf. Sect.~\ref{sec:2.3}).

In that regard, the $r\to\infty$ can be related to setting unit sphere
subtracted by a spherical cap except the North pole
(namely, the partial Riemann sphere), so as to cover
$\mathbb{C}_{\ast(-1/2)}\cup\{\infty\}$.
Here,
$\mathbb{C}_{\ast(m^{\prime})}=\{\mu\bm{\rho}\in\mathbb{C}\mid
0<\mu\|\bm{\rho}\|<\xi_{m^{\prime}}\}$; note that its planar shape
is similar to that for the Cornell regime.
The connection of the complex space to ${\bf R}^{2}$, requisite for
the observation, is established through the North pole from which
to exert stereographic projection of the partial sphere onto
$\mathbb{C}_{\ast(-1/2)}$, and vice versa.
As the normalized plane (c) virtually consists of the two leaves
(sharing the real axis), the extended set is twofold, allowing for
two points at infinity.
Orthogonality of the leaves is to prohibit degeneracy of the points,
to give the distinct $\{+\infty\}$ and $\{-\infty\}$ that anchor
the connection via the real axis.

It turns out, for ${\tilde\lambda}{\hat\sigma}\to\bm{\sigma}/\surd{3}$,
that ${\bf\Lambda}_{0}=({\tilde\lambda}/g_{0})\hat\sigma$ is
rightly reflected in ${\bf S}/(\hbar\surd{3})$.
Herewith, we see that replacement of $\tilde\lambda\to\hbar$
leads to appearance of minimum action; that is, ${\bf S}/\hbar$
turns into small, dimensional operator of ${\bf S}$.
This property is compatible with the characteristic of state vector
and its unitary evolution, regardless of the value of $h$ itself.
Specifically, the $\tilde\lambda=1$ can be interpreted as an
internal unit reflected in quantum bit (``qubit''), i.e.,
the basic unit of quantum information~\cite{schumacher1995}.
The relevant issue is important, which includes responsibility of the internal
mechanics to the quantum entanglement~\cite{schroedinger1935} and
measurements of many-body systems \cite{aspect1982,hensen2015},
but seems to be somewhat out of scope in this paper.

\subsubsection{Concept of Coordinate Self-renormalization}\label{sec:5.2.2}

It is obvious that $\hat\rho$ represents $(n_{d}-1)$-dimensional
projective (tangential) plane of ${\bf r}^{n_{d}}$ space having the
dimensions of $n_{d}=3$, and $\hat\sigma$ does likewise.
In addition, $\xi$ takes on one degree of freedom
that mediates between those two spaces.
Although $\hat\rho$ and $\hat\sigma$ are distinguished from one another,
$\hat\sigma$ in Fig.~\ref{fig:6}(b) just embodies $\hat\rho$ in (a) on the
real axis, so that observer has no way of distinguishing
both the basis vectors for the comparison.
This suggests the isotopic relation of $\hat\sigma$ to $\hat\rho$
by which, after the aforementioned clockwise quarter-turn of $\bm{\rho}$,
$\hat\sigma$ maintains gauge of the real axis, on behalf of $\hat\rho$
that was thereon before the turn.
This instant, $\hat\rho$ is on the imaginary axis, having the magnitude
of $1/\xi$.
That is, as seen in Fig.~\ref{fig:6}(a,b), the turn enforces the
shrinkage of $\hat\rho$, which is undetectable  at this moment.
Especially for the one clockwise turn, the vector shrunk up gets again
on the real axis, to be expressed as $\vD\hat{\rho}=(1/\xi^{4})\hat{\rho}$.
This must be an observable piece.
Meanwhile, for the one turn of $\hat{\rho}$ in Fig.~\ref{fig:6}(c),
$\|\hat{\rho}\|$ is to remain unchanged.
To reconcile the both, we should consider that the apparent
$\hat{\rho}$ having $\|\hat{\rho}\|=1$ results from superposition of
$\vD\hat{\rho}$ on a coordinative vector provisionally normalized
on the real axis in an alternative way.
Expected is that the scalar $\xi$ plays a key r\^{o}le in the primitive
process, referred to as coordinate self-renormalization (CSR), hereafter.

\begin{figure}
\centerline{\includegraphics[trim=165 85 250 135,width=7.4cm,clip]{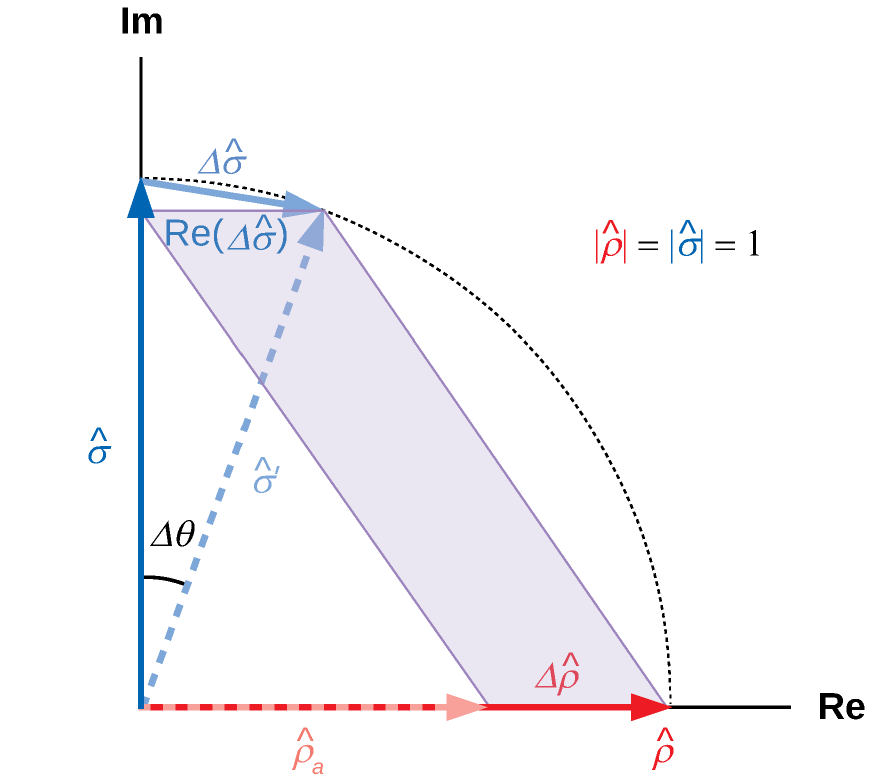}}%
\caption{\label{fig:7}A schematic to explain how to harmonize the
transformation of $\hat{\rho}_{a}\to\hat{\rho}$ with
$\hat{\sigma}\to\hat{\sigma}^{\prime}$.
The dotted curve indicates equator of the Riemann sphere.}
\end{figure}

Concerning the pre-normalization, it is reasonable to recall the critical
value of $\xi$, i.e., $\tilde{\xi}=\mu^{\ast}a$, to give
$\hat{\rho}_{a}=\bm{\rho}/a$ having its magnitude of $\xi/\tilde{\xi}$.
By employing them, one can rewrite $\xi\hat{\rho}$ as
$\tilde{\xi}\hat{\rho}_{a}$ in \eqref{eq:5.6a}.
This prescribes the coordinate setup that enables the interactive
${\bf\Phi}$ to refer to an eigenflow.
Note here that $\|\hat{\rho}_{a}\|<\|\hat{\rho}\|$ because of $\xi<\tilde{\xi}$.
Taking these into account, we make the following transformation
that undertakes the CSR:
\begin{equation}
\hat{\rho}_{a}\,\longrightarrow\,\hat{\rho}
=\hat{\rho}_{a}+\vD\hat{\rho},
\label{eq:5.7}
\end{equation}
\noindent
on $\Re$.
For $\vD\hat{\rho}$ to be observable will require complementation
between $\vD\hat{\rho}$ and $\vD\theta$, in harmony with
the isotopic relation between $\hat\rho$ and $\hat\sigma$.
That is, it is supposed that the transformation of
\begin{equation}
\hat{\sigma}\,\longrightarrow\,\hat{\sigma}^{\prime}
=\hat{\sigma}+\vD\hat{\sigma}
\label{eq:5.8}
\end{equation}
\noindent
is carried out in the way the following relation holds:
\begin{equation}
\vD\hat{\rho}=\Re\left(\vD\hat{\sigma}\right),
\label{eq:5.9}
\end{equation}
\noindent
as displayed in Fig.~\ref{fig:7}.
The CSR can be regarded as the process that transforms the
complex orthogonal superposition of the two isotopic spaces
(as consistent with ${\bf\Psi}_{-}$) into the real superposition,
without violating the condition for the complex space to be the point.

\subsection{Generation of a Numerical Value Comparable to
$\alpha$}\label{sec:5.3}

From \eqref{eq:5.9}, we have
$\|\vD\hat{\rho}\|=\|\hat{\sigma}^{\prime}\|\sin\vD\theta$.
Equating this to the relation of $\|\vD\hat{\rho}\|=1/\xi^{4}$
yields the equation that connects $\xi$ and $\vD\theta$:
\begin{equation}
\xi^{-4}=\sin\vD\theta.
\label{eq:5.10}
\end{equation}
\noindent
Another equation independent of \eqref{eq:5.10} is given by
$\xi\|\hat{\rho}\|=\tilde\xi\|\hat{\rho}_{a}\|$ with
$\|\hat{\rho}_{a}\|=1-\|\vD\hat{\rho}\|$.
We have the expression of
\begin{equation}
\xi =\xi_{-1/2}\left(1-\xi^{-4}\right),
\label{eq:5.11}
\end{equation}
\noindent
where $\tilde\xi$ has been set to the critical value in
$\mathbb{C}_{\ast(-1/2)}$ by which the open set is well defined.
The corresponding $r=a$, denoted as $a_{-1/2}$, specifies
radius of internal cylinder to confine the field
${\bf\Phi}(m=m^{\prime}=-1/2,\kappa>0)$ and its orthogonal
counterpart, as well, by virtue of the consistent coordinate-normalization
of the orthogonal space.
That is, the gauge-fixing $\tilde{\xi}=\xi_{-1/2}$ enables the
eigenflow ${\bf\Phi}(m=m^{\prime}=-1/2,\kappa_{-1/2})$ to regulate
the ${\bf\Psi}_{-}$, properly connecting $\bm{\rho}$ with $\bf R$.
The connection through the infinity, accommodated by the
confinement, is to respond to physical (QED) picture of
coupling the infinitesimal point charge with the photon field.

In order to highlight the physical correspondence, we approximately
solve the set of equations~\eqref{eq:5.10} and \eqref{eq:5.11},
assuming $\vD\theta\ll 1$.
When ignoring $\sim(\vD\theta)^{3}$ and the higher order terms,
\eqref{eq:5.10} reduces to
\begin{equation}
\xi^{-4}\cong\vD\theta,
\label{eq:5.12}
\end{equation}
\noindent
and \eqref{eq:5.11} allows the expression of
\begin{equation}
\xi^{-4}\cong\xi_{-1/2}^{-4}/\bigl(1-4\xi_{-1/2}^{-4}\bigr).
\label{eq:5.13}
\end{equation}
\noindent
Thus, if only the numerical value of $\xi_{-1/2}$ is given
(cf. Table~\ref{tb:1}), one can algebraically calculate the values of
$\xi$ and $\vD\theta$ from \eqref{eq:5.12} and \eqref{eq:5.13}.
By \eqref{eq:5.12}, the correspondence \eqref{eq:5.2} is cast to
$\sim\xi^{-4}\,\leftrightarrow\,\vD\Theta(\beta\to 0)$.
For the theoretical consistency, the scalar quantity $\xi^{-4}$
must indicate a value comparable to $\alpha$~\cite{schwinger1948}.
Indeed, substituting $\xi_{-1/2}^{4}\cong 141$ into \eqref{eq:5.13}
leads to $\xi^{4}\cong 137$ so that the assumption of
$\vD\theta\ll 1$ is valid.
The result suggests that ${\bf\Psi}_{-}$ does not only provide
the geometric representation as compared to the Wigner's
classification, but also be structure, i.e., {\em physical} reality
of the electron.
It is noted that the value of $\xi^{4}$ is unchanged under the
inverse rotation ($m=-m^{\prime}>0$) of the coordinate-rotor with
$\mathrm{sgn}(\kappa)=-1$ (Sect.~\ref{sec:4.3}), regulated by the
common $\mathbb{C}_{\ast(-1/2)}$; this property is responsible for
the $CP$ invariant along with the positron state.

A tick of $\vD\theta\,(\leftarrow\!\!\vD\hat{\rho}>0)=2\pi\delta_{g}$
measures out one cyclotron-period independently of electron velocity.
We thus see that the scalar correction in the $g$-factor
(the magnetic moment anomaly) can be regarded as the modest
appearance of one-dimension of time, which reflects release of the
degree of freedom of $\xi$ through the process properly connecting
the complex $n_{d}$-dimensional space with ${\bf R}^{n_{d}}$ space.
It follows from this that dimensions of space-time are $3n_{d}+1=10$, and
conceivable degrees of freedom add up to $11$.

The number appears to coincide with that required for unification of
the theory having
$\mathrm{SU(3)}\times\mathrm{SU(2)}\times\mathrm{U(1)}$
symmetry and gravitational theory, specifically, that as a
far-reaching consequence in \cite{witten1995}.
Accordingly there is, more or less, a possibility that
the current geometry is linked to the one in the string theory.
However, it should be noted that a significant difference exists between
their background vacua: the field ${\bf\Psi}_{-}$ pervades the realistic
space beyond the point in ${\bf R}^{3}$, and undergoes the cylindrical
confinement in $\xi<\xi_{-1/2}\neq\infty$ when regulated.
As is, the operational methodology to see the internal landscape
through the point (instead of an ingredient having the finite size of
$\sim l_{\mathrm{P}}$) is apparently free from the problem
of compactification incidental to the higher-dimensional theories
extended along Kaluza-Klein theory~\cite{kaluza1921,klein1926},
whereas the stereographic projection corresponds to a class of
Alexandroff one-point compactification of the complex space.
The causal disconnection of the double vacuum, which accompanies
the operation owing to the indeterminate form (Sect.~\ref{sec:2.3}),
is essential for the four-dimensional spatiotemporal setup mentioned
above (also, compatible with the quantum measurements).
This means that the internal mechanics is not subjected to
our space-time; this theory belongs to a category of
background-independent theories in the conventional sense.

Along time-energy correspondence in four vector, the scale of
$\sim\xi^{-4}$ can be translated as origin of electric energy of
the point charge excited in our vacuum.
Then, inequality of $\xi^{-4}>\xi_{-1/2}^{-4}$ is amenable to the
QED vacuum picture in which net charge decreases outward
due to polarization of virtual electron-positron pairs.
Effective thickness of the shielding shell is deemed to reflect
$\vD\xi =\xi_{-1/2}-\xi$.
Denoting by $e_{\mathrm{th}}$ a finite charge that should
theoretically be determined, one has $e^{2}=e_{\mathrm{th}}^{2}/(1-Y_{c})$
with $Y_{c}$ being constant, in natural units~\cite{feynman1961}.
Now we see the remarkable correspondence between the
$e^{2}$-expression and \eqref{eq:5.13};
$Y_{c}$ reflects the quantity of $\sim 4\xi_{-1/2}^{-4}$,
to be expressed as $4e_{\mathrm{th}}^{2}$
approximately.\footnote[4]{One can find ``$e_{\mathrm{th}}^{2}=1/141$''
in a {\em completely speculative} sentence of
Feynman's lecture note~\cite{feynman1961}.}
It is claimed that physical meaning of the charge renormalization,
which copes with photon self-energy, can be ascribed to the
CSR regulated by a rotational eigenvalue for $m^{\prime}<0$.\\

\section{Further Discussions on the Compatibility with the
Standard Model}\label{sec:6}

In order to make the theory self-contained, I provide a concise
explanation of how the internal mechanics could be linked to
an internal representation responsible for the quark and weak charge
and leptonic mass that characterize the electron-based
$\beta$-decay.
The argument is expanded in a way we could track its
compatibility with the standard model.
The manner is reasonable in that the standard model systematically
describes almost all experimental results with
highest accuracy at present.

\subsection{Quark Electric Charge}\label{sec:6.1}

\begin{figure}
\centerline{\includegraphics[trim=135 125 140 165,width=8.5cm,clip]{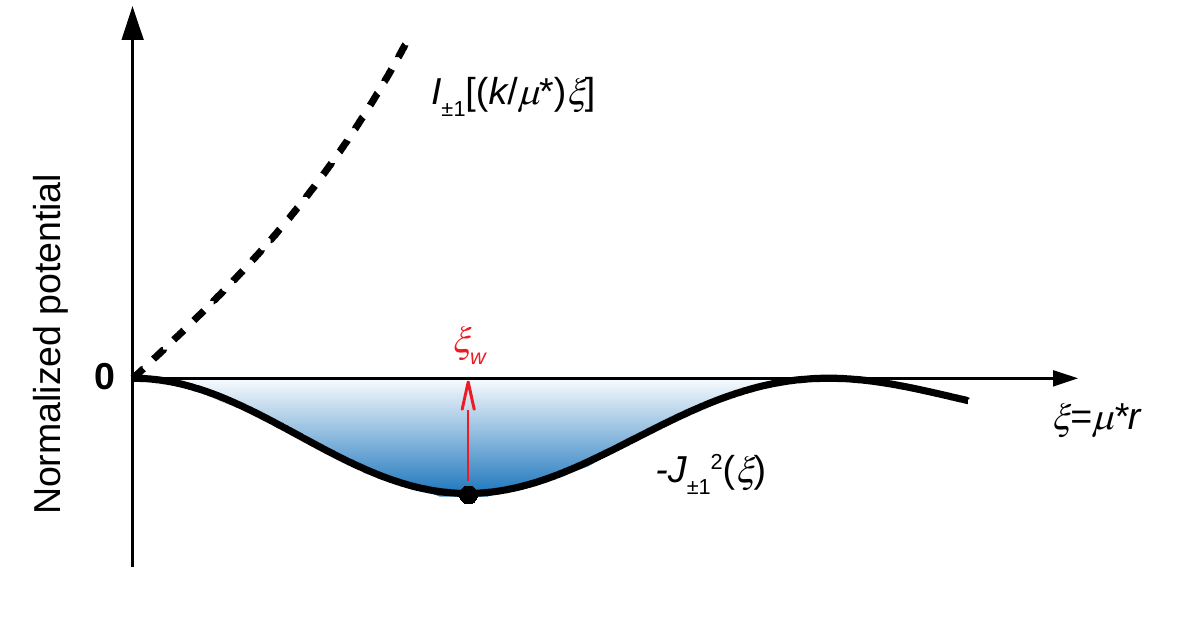}}%
\vspace{-.2cm}
\caption{\label{fig:8}The transformation of the cylindrical functions,
$I_{\pm 1}\to -J_{\pm 1}^{2}$, caused by \eqref{eq:6.1}.
On the horizontal axis, $\xi_{w}$ denotes the value of $\xi$
at which the resulting function as the normalized potential
$U_{\pm}(\xi)/\phi_{\pm 1}^{2}$ (solid curve) gives a minimum value.}
\end{figure}

We attempt to connect the CSR with the representation
potentially reflected in the quark electric charge.
This is a subject to the electroweak interaction in
the closed space confining quarks.
In the internal system in which equation~\eqref{eq:3.4} with $m=\pm 1$
holds, reconsider the region of $r\to\infty$ so that the solution can be
expressed as $\eta(kr)\sim I_{\pm 1}(kr)\gg K_{\pm 1}(kr)$
(Sect.~\ref{sec:4.1}).
This corresponds to the physical situation in which the low-energy
limit is considered far away from the high-energy (perturbative QCD)
region characterized by the asymptotic
freedom~\cite{gross1973,politzer1973}.
The internal $k$-transformation applied to $I_{\pm 1}(kr)$ is
recast to the form of
\begin{equation}
kr\,\longrightarrow\,\varphi\coloneqq -i\xi,
\label{eq:6.1}
\end{equation}
\noindent
with $\xi\coloneqq\mu^{\ast}r$.
It is supposed that, the process in which $W^{\pm ,0}$ and $B^{0}$
particles appear is signified by the transformation of $\eta(kr)$
owing to \eqref{eq:6.1}.
Transformation of the $\eta^{g_{0}}$ norm is written as follows:
\begin{equation}
\eta^{g_{0}}(kr)\,\longrightarrow\,
U_{\pm}(\xi)=-\phi_{\pm 1}^{2}J_{\pm 1}^{2}(\xi).
\label{eq:6.2}
\end{equation}
\noindent
Equation~\eqref{eq:6.2} could represent generation of the
internal potential that acts on $\ell\bar\ell$ pair separated,
in dual relation to the \eqref{eq:3.12} that represents generation of
the external potential of $q\bar{q}$ pair confined.
For convenience, the function transformation of
$I_{\pm 1}(kr)=I_{\pm 1}[(k/\mu^{\ast})\xi]\to-J_{\pm 1}^{2}(\xi)$
is shown in Fig.~\ref{fig:8}.
Since $\phi_{\pm 1}J_{\pm 1}(\xi)$ constitutes $\Phi_{z}$, it is
reasonable to suppose that
${\bf\Phi}(m=m^{\prime}=\pm 1,\mp\kappa_{\pm 1})$ internally
regulates the weak interaction, as the
${\bf\Phi}(m=m^{\prime}=-1/2,\kappa_{-1/2})$
regulates the electromagnetic interaction.
Henceforth, the eigenflows for which $m=m^{\prime}$ are denoted as
${\bf\Phi}_{m}$.

We focus on the $\beta^{-}$-decay reaction, decomposed into
$d_{(-1/2)}\longrightarrow u_{(1/2)}+W_{(-1)}^{-}$
and $W_{(-1)}^{-}+\nu_{e(1/2)}\longrightarrow e_{(-1/2)}^{-}$.
Here, the values in the brackets indicate the third component of
the weak isospin, $T_{3}$.
Along the relation of ${\bf\Phi}_{-1/2}$ with $e_{(-1/2)}^{-}$,
we make the correspondence between $m$ and $T_{3}$ for
particle species endowed with negative charge and negative $T_{3}$.
The Yukawa interaction is rewritten in the form
\begin{equation}
\bigl[d_{\mathrm L}+\left(\bar u\right)_{\mathrm R}\bigr]\,
\longrightarrow\,W^{-},\mspace{30mu}
W^{-}\,\longrightarrow\,e_{\mathrm L}^{-}
+\left(\bar{\nu}_{e}\right)_{\mathrm R}.
\label{eq:6.3}
\end{equation}
\noindent
Here, the square bracket emphasizes the fact that the $d\bar{u}$ pair
is confined in the closed space of pion interior.

Now we reconsider the charged current interaction of $d_{\mathrm L}$
(undergoing rotational motion in the closed space) with $W^{-}$.
We read that the left-handed $d_{\mathrm L}$, involved in the interaction
along with $e_{\mathrm L}^{-}$, goes for the CSR owing to \eqref{eq:5.6a}
with $\mathrm{sgn}(\kappa)=+1$, as basic internal mechanics of $T_{3}=-1/2$
[$(\bar{d})_{\mathrm R}$ and $e_{\mathrm R}^{+}$, \eqref{eq:5.6a}
with $\mathrm{sgn}(\kappa)=-1$, as that of $T_{3}=1/2$].
It is seen that for rotational motion of $e^{-}$, the CSR referring to
$\mathbb{C}_{\ast(-1/2)}$ sets up a proper rotational coordinate
in the ${\bf R}^{3}$ space that allows the orbit.
This setup is none other than development of electromagnetic potential.
In this view, is expected that the CSR referring to $\mathbb{C}_{\ast(-1)}$
contributes to setting up a proper coordinate, i.e., developing the
interaction potential, in the closed space of $J^{P}=0^{-}$ particle.
In particular, the gauge-fixing $\tilde{\xi}=\xi_{-1}$ must be involved in
determining negative charge of the $d$ quark, in the same way
as the determination of $-e$ in which $\tilde{\xi}=\xi_{-1/2}$ is involved.
In this regime, the $k\to 0$ as a premise of \eqref{eq:5.6a} is
supposed to capture the lowest energy excitation of the pion,
instead of relativistic excitation of the constituent quarks.

From this notion follows: the $\xi^{-4}$-scale of \eqref{eq:5.13} with
$\xi_{-1}$ replacing $\xi_{-1/2}$ must be responsible for square of
the $d$ quark charge $-e_{q}$.
This replacement owes to base coupling between
${\bf\Phi}_{-1/2}$ and ${\bf\Phi}_{-1}$ via $\xi$.
As a consequence, we have the following expression compared to
$\sqrt{e_{q}^{2}/e^{2}}=1/3$~\cite{gellmann1964}:
\begin{equation}
\sqrt{(\xi_{-1/2}^{4}-4)/(\xi_{-1}^{4}-4)}\cong 0.333,
\label{eq:6.4}
\end{equation}
\noindent
where use has been made of $\xi_{-1}^{4}\cong 1240$ (Table~\ref{tb:3}).
The outcome indicates the overall consistency with the quark model
relying on $\mathrm{SU(3)}$ symmetry, supporting the relevance of
the potential theory that has preliminarily been developed in
Sect.~\ref{sec:3}.

\subsection{Vacuum and Mass Generation}\label{sec:6.2}

We further examine the transformation of \eqref{eq:6.2}, in terms of the
mass change $m_{W}\to m_{\ell,\bar{\ell}}$ in the second equation of
\eqref{eq:6.3}.
By the transformation, the value of $\xi$ at which the potential indicates
the minimum changes from $0$ to $\xi_{w}\neq 0$, as seen in Fig.~\ref{fig:8}.
The value of $\xi_{w}$ characterizes the quantity $\mu^{\ast}$ included
in ${\bf\Phi}_{-1}$, so that $\xi_{w}$ must be reflected in $m_{W}$.

For real $\xi$, expansion of $J_{\pm 1}^{2}(\xi)$ can be expressed as
a function of $\xi^{2}=\varphi^{\ast}\varphi$, where the asterisk
signifies the complex conjugate.
Accordingly, we have
\begin{equation}
U_{\pm}(\varphi^{\ast}\varphi)=
-\frac{\phi_{\pm 1}^{2}}{4}\left[\varphi^{\ast}\varphi
-\frac{1}{4}(\varphi^{\ast}\varphi)^{2}
+\mathcal{O}\left((\varphi^{\ast}\varphi)^{3}\right)\right].
\label{eq:6.6}
\end{equation}
\noindent
For a comparison, we shall call the standard vacuum
model~\cite{higgs1964}, in which one assumes condensate of
complex scalar field (denoted as $\varphi^{\prime}$).
Potential energy(-density) of the self-interacting $\varphi^{\prime}$
is written in the approximate form of
$U(\varphi^{\prime\ast}\varphi^{\prime})\cong
\mu^{2}\varphi^{\prime\ast}\varphi^{\prime}
+\lambda\left(\varphi^{\prime\ast}\varphi^{\prime}\right)^{2}$,
where $\mu^{2}$ and $\lambda$ are constants.
Vacuum state is represented by the expectation value of
$\varphi^{\prime}$, such as $\left<\varphi^{\prime}\right>_{0}=0$
for $\mu^{2}>0$: the value of
$\sqrt{\varphi^{\prime\ast}\varphi^{\prime}}$
at which $U$ indicates the minimum.
As for $\mu^{2}<0$, it is found that $U$ indicates the minimum at
$\varphi^{\prime\ast}\varphi^{\prime}=-\mu^{2}/2\lambda$, to give
$\left<\varphi^{\prime}\right>_{0}=\sqrt{-\mu^{2}/2\lambda}$,
which linearly couples with $m_{W}$.
Quantum excitation of the field can be expressed as
$\varphi^{\prime}=\left<\varphi^{\prime}\right>_{0}+H/\surd{2}$,
where $H$ the Higgs boson.
Apparently, the analytic function of $U_{\pm}(\varphi^{\ast}\varphi)$ is
compatible with $U(\varphi^{\prime\ast}\varphi^{\prime})$ for $\mu^{2}<0$.
Provided that $\varphi^{\prime\ast}\varphi^{\prime}$ in our space-time
reflects $\varphi^{\ast}\varphi$, thus, we have
\begin{equation}
\sqrt{\varphi^{\ast}\varphi}=\xi_{w}\,\longleftrightarrow\,
\left<\varphi^{\prime}\right>_{0},
\label{eq:6.7}
\end{equation}
\noindent
which supports the relation of $\xi_{w}\propto m_{W}$.
The transformation \eqref{eq:6.1} and the definite onset of $\hat{\bf z}$
are found to be requisite for generating $\xi_{w}$.

In connection with the chiral asymmetry of ${\bf\Phi}_{\pm\|m\|}$, recall
that $(\bar{\nu}_{e})_{\mathrm R}$ is to internally refer to ${\bf\Phi}_{1/2}$.
The natural extension leads to the idea that the spin $1$ neutral boson
that can interact with right-handed particles, namely $B^{0}$,
refers to ${\bf\Phi}_{1}$.
For the eigenvalue, in keeping with what $\xi_{-1}$ contributes to
determining $e_{q}$ of the $J^{P}=0^{-}$ particle, we conjecture
that its counterpart, $\xi_{1}$, indicates mass of $0^{+}$ particle,
specifically, the Higgs boson mass $m_{\mathrm H}$.
The point is that, provided a pair of ${\bf\Phi}_{\pm 1}$ shares
the same base coordinate with the common variable $a$,
the ratio $\xi_{1}/\xi_{w}$ must be reflected in mass ratio of
$m_{\mathrm H}/m_{W}$.
The specific values of $\xi_{1}\cong 2.86$ and $\xi_{w}\cong 1.84$
give $\xi_{1}/\xi_{w}\cong 1.55$.
This indeed accords with $m_{\mathrm H}/m_{W}\cong 1.55$ for the
experimental values of $m_{W}\cong 80.4\,\mathrm{GeV}/c^{2}$\,
\cite{workman2022,aaltonen2022} and
$m_{\mathrm H}\cong 125\,{\mathrm{GeV}/c^{2}}$~\cite{aad2015}.

We see that $\xi_{w}=\sqrt{\varphi^{\ast}\varphi}\leftarrow\xi_{1}$
is responsible for $m_{W}\leftarrow m_{\mathrm H}$.
Then, the $m_{W}\to m_{\ell,\bar{\ell}}$ is expected to
reflect $\xi_{w}\to\xi_{\mp 1/2}$, which cooperates with the
system transformation of $\|m\|=1\to m=\mp1/2$,
in which the confinement radius of ${\bf\Psi}_{\mp}$ is set to as
$r=a\to a_{\mp 1/2}$.
The base coupling $\sqrt{\varphi^{\ast}\varphi}=\xi_{\mp 1/2}$
is necessary for the regulated ${\bf\Psi}_{\mp}$
to properly couple with $U_{\mp}(\varphi^{\ast}\varphi)$.
In the context, $\xi_{1}\to\xi_{\mp 1/2}$ symbolically represents a process
in which the rest mass $m_{\ell,\bar{\ell}}$ is imprinted in our vacuum.
When comparing $\xi=\xi_{-1/2}-\vD\xi\to\xi_{-1/2}$ to mass
renormalization of a charged lepton, $\xi\rightleftarrows\xi_{-1/2}$
signifies self-complementation between the
mass and charge renormalization.

We are tempted to relate the imprint of $m_{\ell,\bar{\ell}}$ to
excitation of the $\varphi$-field in $U_{\mp}$.
The excitation energies denoted as $\omega_{-1/2}$ and $\omega_{1/2}$
are given by
\begin{equation}
\omega_{-1/2}=U_{-}(\xi_{-1/2})-U_{-}(\xi_{w}),\qquad
\omega_{1/2}=U_{+}(\xi_{1/2})-U_{+}(\xi_{w}).
\label{eq:6.8}
\end{equation}
\noindent
It is noted that if $\phi_{1}^{2}/\phi_{-1}^{2}\sim\mathcal{O}(1)$,
we have the relation of $\omega_{1/2}\ll\omega_{-1/2}$, because of
$U_{+}(\xi_{w})\approx U_{+}(\xi_{1/2})$ due to $\xi_{w}\sim\xi_{1/2}$.
Equation~\eqref{eq:6.8} can be compared to the internal mass
of ${\bf\Psi}_{\mp}$ analogous to meson mass.
In this aspect, we invoke the Gell-Mann-Oakes-Renner relation
indicating that $m_{\pi}^{2}$ is proportional to mass of
the constituent quarks~\cite{gellmann1968}.
Combining these, we read ratio of $m_{\bar\ell}/m_{\ell}$ as reflection
of $(\omega_{1/2}/\omega_{-1/2})^{2}\cong 3.53\times10^{-6}
(\phi_{1}^{2}/\phi_{-1}^{2})^{2}$, and also,
$a_{1/2}/a_{-1/2}=(\xi_{1/2}/\xi_{-1/2})(\omega_{-1/2}/\omega_{1/2})^{2}$.
These values are estimated below.

\subsection{The Electroweak Coupling}\label{sec:6.3}

Reminding the form of the momentum operator that suffices for
$\mathrm{SU(2)}\times\mathrm{U(1)}$ gauge invariance,
we attempt at extending the indication of
$\phi_{-1/2}^{2}\to\alpha$ (Sect.~\ref{sec:4.1}):
natural is to simply suppose $\phi_{-1}^{2}\to\vg^{2}/q_{\mathrm{P}}^{2}$
and $\phi_{1}^{2}\to \vg^{\prime 2}/q_{\mathrm{P}}^{2}$, where
$\vg$ ($\vg^{\prime}$) is the coupling strength between the
$W$ ($B^{0}$) boson and the weak isospin (weak hypercharge)
of left-handed (both-handed) particles.
Therewith, the idea is that mode coupling among ${\bf\Phi}_{m}$, with
$\phi_{m}$ being its strength, describes the electroweak coupling.
At this juncture, we recall the orthogonal decomposition of
$\kappa_{m(=m^{\prime})}$-spectrum: $(\xi_{m},k_{m})$.
As we have seen that $m_{\mathrm H}\to m_{\ell,\bar{\ell}}$ goes for
$\xi_{m}$ with $m=1\rightarrow\mp 1/2$, the $\tilde\xi$-space is
exhausted by coupling with matter fields.
It is, therefore, reasonable to consider that the $\tilde k$-space
left over is used for the concerned mode coupling, thereby
yielding a representation of gauge field unification.
Synchronizing with the mass pointer, $(\vg,\vg^{\prime})\to e$ as
charge pointer is to go for $k_{m}$ with $m=\mp 1\rightarrow -1/2$.

A vital clue to the relation between $\phi_{m}$ and $k_{m}$ is found in
the coupling of cylindrical functions of order $m=\pm 1$
(Sect.~\ref{sec:3.1}).
An attention should be paid to the ratio of
$c_{-1}^{2}/c_{1}^{2}\sim\tilde{k}^{4}$, which plays a critical r\^{o}le
in the quark confinement.
Altogether, the confinement of ${\bf\Psi}_{-}$ inside the cylinder of
${\bf\Phi}_{-1/2}$, which is compared to perfectly conducting wall,
resembles the confinement of the color electric field
as seen in the dual superconductor picture~\cite{nambu1974}.
Hence, along the $\tilde{k}$-scaling form, we describe
the multi-mode coupling among ${\bf\Phi}_{m}$.
We here introduce the hyper coupling constant $\phi_{p}^{2}$
[$\sim(\xi_{-1/2}k_{-1/2})^{-4}$], and hypothesize that
the flows of ${\bf\Phi}_{m}$ are self-adjusted in the way their
intensities satisfy $\phi_{m}^{2}/\phi_{p}^{2}=k_{m}^{4}$ each.
Then, we have the chain rule for $m=-1/2$, $-1$, and $1$,
expressed as follows:
\begin{equation}
\phi_{p}^{2}=\frac{\phi_{-1/2}^{2}}{k_{-1/2}^{4}}
=\frac{\phi_{-1}^{2}}{k_{-1}^{4}}=\frac{\phi_{1}^{2}}{k_{1}^{4}}.
\label{eq:6.9}
\end{equation}
\noindent
In particular, we obtain the ratio evaluated as
\begin{equation}
\phi_{-1/2}^{2}/\phi_{-1}^{2}=
\left(k_{-1/2}/k_{-1}\right)^{4}\cong 0.236.
\label{eq:6.10}
\end{equation}
This ratio is to be reflected in
$e^{2}/\vg^{2}=\sin^{2}\theta_{\mathrm W}$ in a low energy region,
where $\theta_{\mathrm W}$ is the weak mixing angle~\cite{weinberg1967}.
Indeed, the value of equation~\eqref{eq:6.10} agrees with the
experimental data (marginally smaller than about $0.239$ for the
standard model) from atomic parity violation in
sub-$10^{-2}\,\mathrm{GeV}$ range (see figure~10.2 in
\cite{workman2022}, and references therein).
Also, $\phi_{1}/\phi_{-1/2}=(k_{1}/k_{-1/2})^{2}\cong 1.14$
is compared with $\vg^{\prime}/e$, to be almost consistent with the
corresponding ratio of $m_{Z}/m_{W}$, where $m_{Z}$ the $Z$ boson mass.

The numerical value of $m_{\bar\ell}/m_{\ell}$ works out, by letting
$\phi_{1}^{2}/\phi_{-1}^{2}=(k_{1}/k_{-1})^{4}$ in the foregoing expression
of $(\omega_{1/2}/\omega_{-1/2})^{2}$, at $3.41\times 10^{-7}$,
and the one of $a_{1/2}/a_{-1/2}$, at $1.61\times 10^{8}$.
Making the former ratio correspond to mass ratio of
$\bar{\nu}_{e}$ to $e^{-}$, the $\bar{\nu}_{e}$ mass is estimated to be
$m_{\bar{\nu}_{e}}\cong 174\,\mathrm{meV}/c^{2}$, as consistent with
the upper limit from direct neutrino mass experiment~\cite{aker2022}.
The experiment being in progress, or future planned
ones~\cite{esfahani2017,betti2019}, could verify the
proposed scenario.\\

\section{Conclusions}\label{sec:7}

In conclusion, I have proposed the internal geometric structure of
point particles, which could capture the coupling with gauge fields
and generate their intrinsic attributes.
Infinitesimal realistic space with higher-dimensions has been
supposed in a singular region of the point particles.
I have suggested plasma dielectric (massive photon) picture of
transformation between the internal space and our conventional space,
as well as, magnetohydrodynamic (magnetic reconnection) picture of
topological transformation of internal harmonic function.
From the scalar-to-vector transformation, we have derived
the rotational field capable of representing the Dirac's algebra.
The vector field, which satisfies rotational eigenvalue equation,
resides in the extra-dimensional space; it is updated recurrence of
the Chandrasekhar-Kendall solution for Gromeka-Beltrami flow.
The eigenflow that could be reflected in electromagnetic
susceptibility has been investigated in details.
A helical module representation of $\beta^{-}$-decay products
has been provided in comparison with the Wigner's representation.

From the rotational eigenvalue equation, we have drawn a coordinate-rotor
accommodated by the isotopic relation between complex orthogonal spaces.
By turning the rotor, we have seen a mechanical sequence to maintain
basis vectors in the spaces, as responsible for cyclotron motion of
electron with its spin precession.
It has been shown that the CSR (coordinate self-renormalization)
process regulated by a specific rotational eigenvalue could be
compared to charge renormalization.
Within this framework, the CSR is found to, also, take on creating
temporal degree of freedom, generating a numerical value comparable
to the electromagnetic coupling constant in low-energy limit.
And, the new picture is obtained in which electrons embody reality
of the geometric structure to ensure the scalar-intervening
connection between our space and the infinitesimal space
pertaining to another background.

Chiral asymmetry of the eigenflows has been connected
with parity violation in weak interaction.
A consequence is that the left-handed eigenflows
${\bf\Phi}_{-1/2,\,-1}$ are involved in determining negative charge
of lepton and quark, and multi-mode coupling of
${\bf\Phi}_{-1/2,\,\mp 1}$ sustains the electroweak coupling,
generating a numerical value comparable to the Weinberg angle
measured at low energy.
The theory suggests an internal landscape of
$\mathrm{SU(3)}\times\mathrm{SU(2)}\times\mathrm{U(1)}$
Yang-Mills fields plus Higgs vacuum, though it is speculative,
and has only been compared with the partial structure of the
quantum field theory at the moment.
It is also notified that the following questions remain open:
whether or not/how
(1) this framework could respond to the higher energy including
the so-called running and higher generation of leptons (related to
higher order loop corrections of the anomalous magnetic moment),
(2) the geometry could be connected with that of the other
hypothetical theories, particularly, the string theory,
(3) the CSR, with the general coordinate transformation,
(4) the operational transformation, with the quantum measurements,
and so forth.

In any case, the obtained results tentatively support the common
picture of helical plasmas for the internal structure of point particles.
Interestingly, the theoretical prototype seems to reveal some
characteristics a parameter-free framework (if any, away from the
anthropic principle~\cite{dicke1961,carter1974}) is likely to have.
Regarding this, the extended scenario of determining the neutrino mass
may be testable by laboratory experiments.
I hope this work adds to particle physics in future, bringing a fresh
perspective on structure of the Universe.\\

\backmatter

\noindent
{\bf Acknowledgements}\quad This version of the article has been
accepted for publication, after peer review (when applicable)
but is not the Version of Record and does not reflect
post-acceptance improvements, or any corrections.
The Version of Record is available online at:
\url{https://doi.org/10.1007/s10773-024-05717-5}.
Use of this Accepted Version is subject to the publisher’s
Accepted Manuscript terms of use
\url{https://www.springernature.com/gp/open-research/policies/accepted-
manuscript-terms}\\

\noindent
{\bf Author Contributions}\quad M.H. is the solo author of the manuscript.\\

\noindent
{\bf Funding}\quad No funds, grants, or other support was received.\\

\noindent
{\bf Data Availability}\quad All data generated or analysed during this study
are included in this published article.\\

\noindent
{\large\bf Declarations}\\

\noindent
{\bf Competing Interests}\quad The author declares no competing interests.\\

\end{document}